\documentclass[%
 aip,
 amsmath,amssymb,
 reprint,%
]{revtex4-1}

\usepackage{graphicx}
\usepackage{dcolumn}
\usepackage{bm}

\usepackage[utf8]{inputenc}
\usepackage[T1]{fontenc}
\usepackage{mathptmx}
\usepackage{etoolbox}

\usepackage{multirow}

\makeatletter
\def\@email#1#2{%
 \endgroup
 \patchcmd{\titleblock@produce}
  {\frontmatter@RRAPformat}
  {\frontmatter@RRAPformat{\produce@RRAP{*#1\href{mailto:#2}{#2}}}\frontmatter@RRAPformat}
  {}{}
}%
\makeatother

\usepackage{xcolor}
\newcommand{\com}[1]{{\color{black}#1}}

\begin{document}

\preprint{AIP/123-QED}

\title[Asymmetry of wetting and de-wetting on high-friction surfaces originates from the same molecular physics]{Asymmetry of wetting and de-wetting on high-friction surfaces originates from the same molecular physics}
\author{M. Pellegrino}
\email{micpel@kth.se}
\affiliation{Swedish e-Science Research Centre, Science for Life Laboratory,
Department of Applied Physics KTH, 100 44 Stockholm, Sweden}
\author{B. Hess}%
\email{hess@kth.se}
\affiliation{Swedish e-Science Research Centre, Science for Life Laboratory, Department of Applied Physics KTH, 100 44 Stockholm, Sweden}

\date{\today}

\begin{abstract}
    The motion of three-phase contact lines is one of the most relevant research topics of micro- and nano-fluidics. According to many hydrodynamic and molecular models, the dynamics of contact lines is assumed overdamped and dominated by localised liquid-solid friction, entailing the existence of a mobility relation between contact line speed and microscopic contact angle. We present and discuss a set of non-equilibrium atomistic Molecular Dynamics simulations of water nanodroplets spreading on or confined between silica-like walls, showing the existence of the aforementioned relation and its invariance under wetting modes (`spontaneous' or `forced'). Upon changing the wettability of the walls, it has been noticed that more hydrophilic substrates are easier to wet rather than de-wet; we show how this asymmetry can be automatically captured by a contact line friction model that accounts for the molecular transport between liquid layers. A simple examination of the order and orientation of near-contact-line water molecules corroborates the physical foundation of the model. \com{Furthermore, we present a way to utilize the framework of multicomponent molecular kinetic theory to analyze molecular contributions to the motion of contact lines}. Lastly, we propose an approach to discriminate between contact line friction models which overcomes the limitations of experimental resolution. This work constitutes a stepping stone towards demystifying wetting dynamics on high-friction hydrophilic substrates and underlines the relevance of contact line friction in modelling the motion of three-phase contact lines.
\end{abstract}

\maketitle

\section{Introduction}

The motion of fluid-fluid interfaces over solid substrates is a common occurrence in processes both related to the natural world and to industry \cite{bonn2009wettingspreading, kumar2015liquid}. The first attempts to describe the dynamics of a contact line \com{(c.l.)} using the tools of continuous fluid mechanics encountered an obstacle represented by the fact that no-slip boundary conditions entail infinite stresses at the substrate, would the contact line move \cite{huh1970hydrodynamic}. Chasing this mathematical paradox, many different attempts to fully regularise the problem have emerged \cite{Voinov1976model, cox1986viscous, petrov1992combined, shikhmurzaev1997ift}. Among those, one of the most popular is the contact line friction model which stems from the Molecular Kinetic Theory (MKT) by Blake and Haynes \cite{blake1969mkt}: under the assumptions of negligible inertial effects and total dissipation of the fluid's kinetic energy due to friction localised in the contact line region, one obtains a direct mobility relation between contact line speed and microscopic dynamic contact angle: $U_{cl}(t)=f\big(\theta(t)\big)$.

While the existence of a microscopic dynamic contact angle has now been widely accepted by the fluid dynamic and wetting communities, the existence and the characterization of the speed-angle relation still remains controversial and criticised \cite{shikhmurzaev2020litmus}. One of the key arguments against the employment of mobility models is the difficulty to reconcile results obtained under `spontaneous' wetting, where contact lines are driven by capillary action due to out-of-equilibrium initial conditions, and `forced' wetting, where contact lines are driven by mechanical action \cite{karim2016forcedspontaneous}. Proving or disproving contact line models via physical experiments is inherently arduous due to the optical resolution required to resolve the interface curvature in the contact line region. Recently, studies involving optical microscopy and atomic force microscopy have managed to infer the value of microscopic dynamic contact angles \cite{deng2016experimental, lhermerout2016rheometer, eriksson2019meniscus}. However, finely controlling flow conditions and surface wettability in experiments still remains challenging. From this perspective, atomistic Molecular Dynamics (MD) simulations can be used in lieu of experiments to access nanoscopic length and time scales and obtain almost unrestrained control over fluid and surface features. Both spontaneous and forced wetting modes are easily reproducible in MD by either placing a droplet on a surface and changing the surface wettability so that the initial contact angle does not match its equilibrium value (spontaneous), or by confining it between two solid walls and shear them in opposite directions to reproduce a two-phase Couette flow (forced).

Despite MD gaining popularity in studying contact line problems, most research is focused on rather simple cases, such as Lennard-Jones fluids wetting Lennard-Jones crystals. Simulating more complex fluid/surface combinations usually demands more elaborate molecular models and more computational resources. Despite technical challenges, recent works have shown how non-equilibrium MD (NEMD) simulations of water droplets confined between silica-like walls can be effectively employed to characterise continuous simulation techniques `equipped' with dynamic wetting boundary conditions that do not rely on imposing unphysically large slip lengths \cite{lacis2020johansson, lacis2021pellegrino}. Furthermore, upon directly inspecting the dynamic contact angle from NEMD simulations and correlating it with the contact line velocity, it was discovered that advancing and receding contact lines experience different contact line friction; this phenomenon has not been observed for Lennard-Jones substrates \cite{toledano2020hidden}. In particular, the contact line friction on hydrophobic substrates is smaller for receding contact lines and larger for advancing ones, while hydrophilic substrates show the opposite behaviour \com{\cite{lacis2021pellegrino}}. Interpreting the motion of advancing c.l. as `wetting' and the one of receding c.l. as `de-wetting', \textit{hydrophilic substrates are easier to wet rather than de-wet}, and vice-versa. This seemingly intuitive observation fueled our interest to further investigate the asymmetry between advancing and receding contact lines.

MKT predicts an odd functional relationship between contact line speed and uncompensated Young's stress, i.e. it predicts the same contact line speed for the same contact angle deviation, with the sign corresponding to whether the contact line is advancing or receding; hence it cannot capture contact line friction asymmetry naturally. A recent model inspired by the physico-chemistry of water on high-friction surfaces predicts the contact line coefficient to change depending on the \com{absolute} value of the dynamic contact angle \cite{johansson2019friction} and thus represents a good candidate to automatically capture contact line asymmetry. While MKT considers only molecular motion in the wall-nearest liquid layer responsible for the motion of the contact line, this new model adds the contribution of transport from the liquid layer above (in the form of `rolling' motion). It is well known that hydrophilic polar surfaces modify the hydration structure of interfacial water \cite{giovambattista2007polarity}; the importance of considering two fluid layers instead of one is corroborated by simple observations on the conformation of water molecules in the contact line region. \com{As an additional effort towards showing the importance of hydration structure, we propose a method to systematically connect molecular conformation to molecular motion in order to extract kinematic information regarding the processes that advance or recede contact lines. The method involves analysing molecular transitions with the toolbox of Markov state models.}

The purposes of this work are to advance the physical understanding of contact line friction and to show how Molecular Dynamics can be employed to characterise the mobility of three-phase contact lines. These newfound insights can lead to a better modelling of the boundary conditions of continuous fluid mechanics. In order to ensure the reproducibility of our work and to share a set of MD benchmarks useful to further validate contact line models, the output of our MD simulations will be provided in openly accessible databases.

The paper is organised as follows: \com{in section \ref{sec:line-friction} we illustrate the mathematical model of contact line friction. In section \ref{sec:mod-met} we illustrate the NEMD and the analysis techniques employed in this work. In section \ref{sec:res-dis} we illustrate and briefly discuss the main results of molecular simulations. In section \ref{sec:kinematics} we present the study of molecular kinematics of contact lines. In section \ref{sec:disc} we further discuss the implications of our results and finally in section \ref{sec:conc} we draw the conclusions about the impact of the present work and how to proceed forward.}



\section{Contact line friction} \label{sec:line-friction}

\begin{figure}[htbp]
    \centering
    \includegraphics[width=0.30\textwidth]{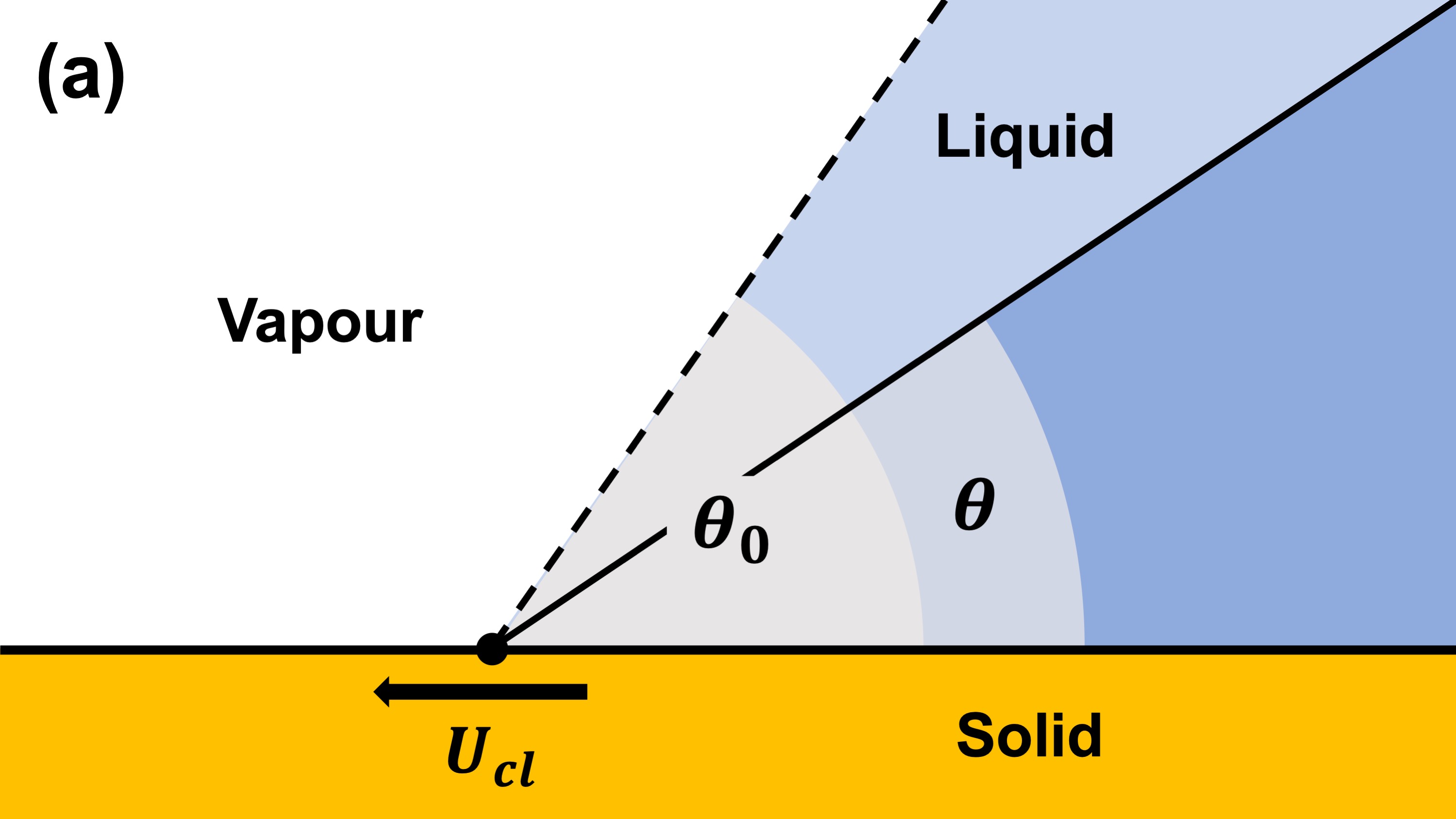}\\
    \vspace{0.5cm}
    \includegraphics[width=0.30\textwidth]{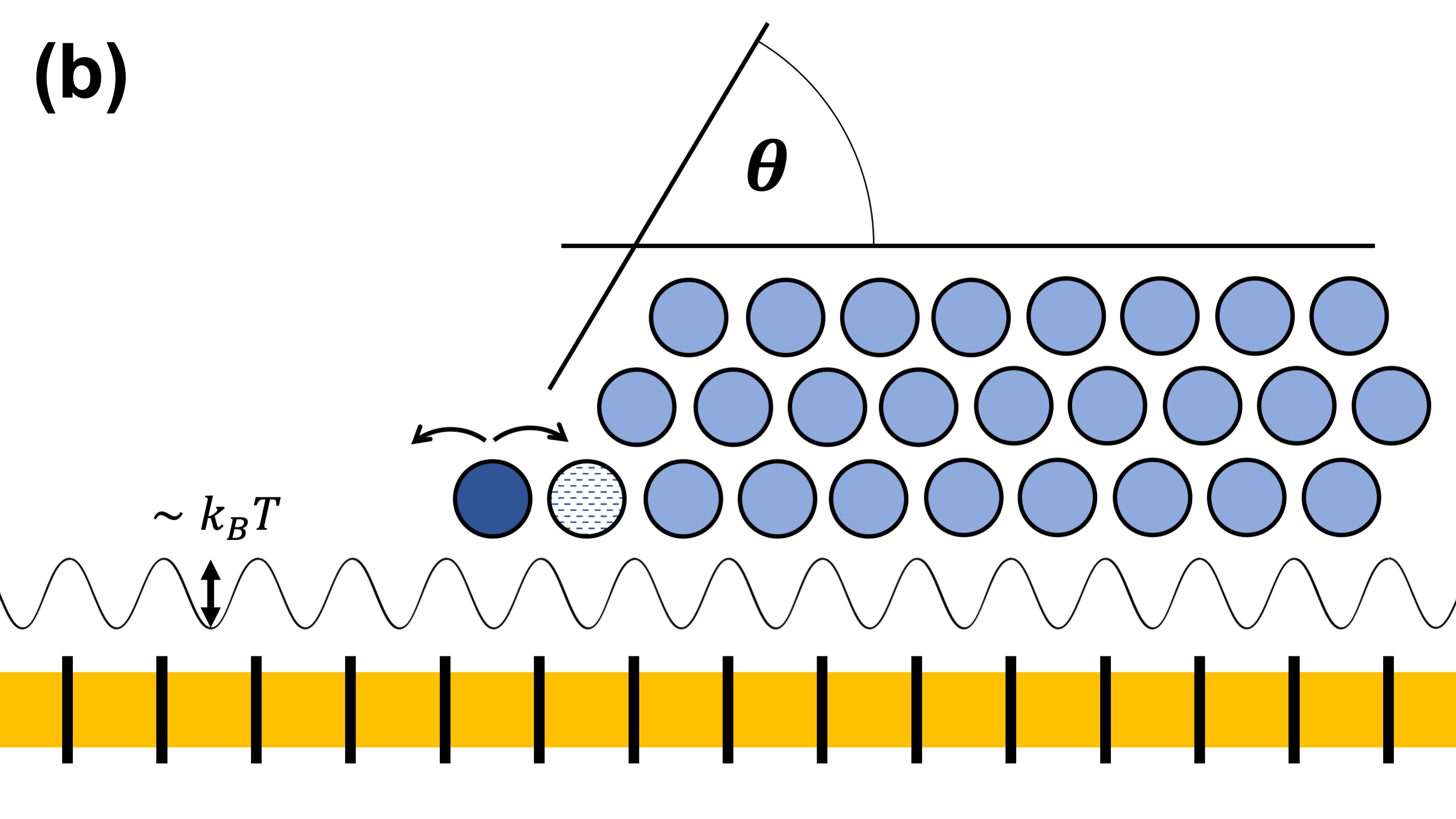}
    \caption{(a) Illustration of a two-dimensional moving contact line (receding) with the observables of interest. (b) Illustration of the hopping mechanism of MKT; the thin wavy solid line represents the substrate's potential energy surface.}
    \label{fig:contact-line}
\end{figure}

We consider the case of a 2-dimensional droplet or meniscus. The wetted surface is atomistically flat and chemically homogeneous. Under such conditions, the equilibrium wetting configuration of a droplet or a meniscus can be uniquely identified by the Young-Dupr\'{e} contact angle $\theta_0$ between the liquid-vapour interface and the solid surface (Fig.~\ref{fig:contact-line}a).

The contact line friction model assumes that when a contact line moves with steady velocity $U_{cl}$ the microscopic contact angle $\theta$ deviates from its equilibrium value, causing an uncompensated Young's stress, quantifiable as $F_Y=\gamma\big(\cos\theta_0-\cos\theta\big)$, being $\gamma$ the liquid-vapour surface tension. Molecular Kinetic Theory describes the motion of contact lines as a sequence of thermally activated molecular `jumps': liquid molecules will preferably stay close to the minima of the potential energy surface of the solid substrate and rattle until they have enough thermal energy to escape, jump, and get adsorbed in another local minimum (Fig.~ \ref{fig:contact-line}b). The mobility of contact lines is therefore determined by the `corrugation' of the surface energy landscape rather than the equilibrium contact angle itself \cite{anhho2011slip}. Furthermore, thermally activated molecular hopping is inherently dissipative. Hence MKT predicts a direct mobility relation between $U_{cl}$ and $F_Y$, which after linearizing for small angle deviations reads:
\begin{equation}    \label{eq:linear-mkt}
    \mu_f U_{cl} = F_Y =  \gamma (\cos\theta_0-\cos\theta\big) \; ,
\end{equation}
where $\mu_f$ is the contact line friction parameter. Although contact line friction depends ultimately on the details of the potential energy surface, it can be correlated to the equilibrium contact angle if the surface topography and chemical composition are kept fixed \cite{lacis2021pellegrino,toledano2020hidden}.

\begin{figure}[htbp]
    \centering
    \includegraphics[width=0.30\textwidth]{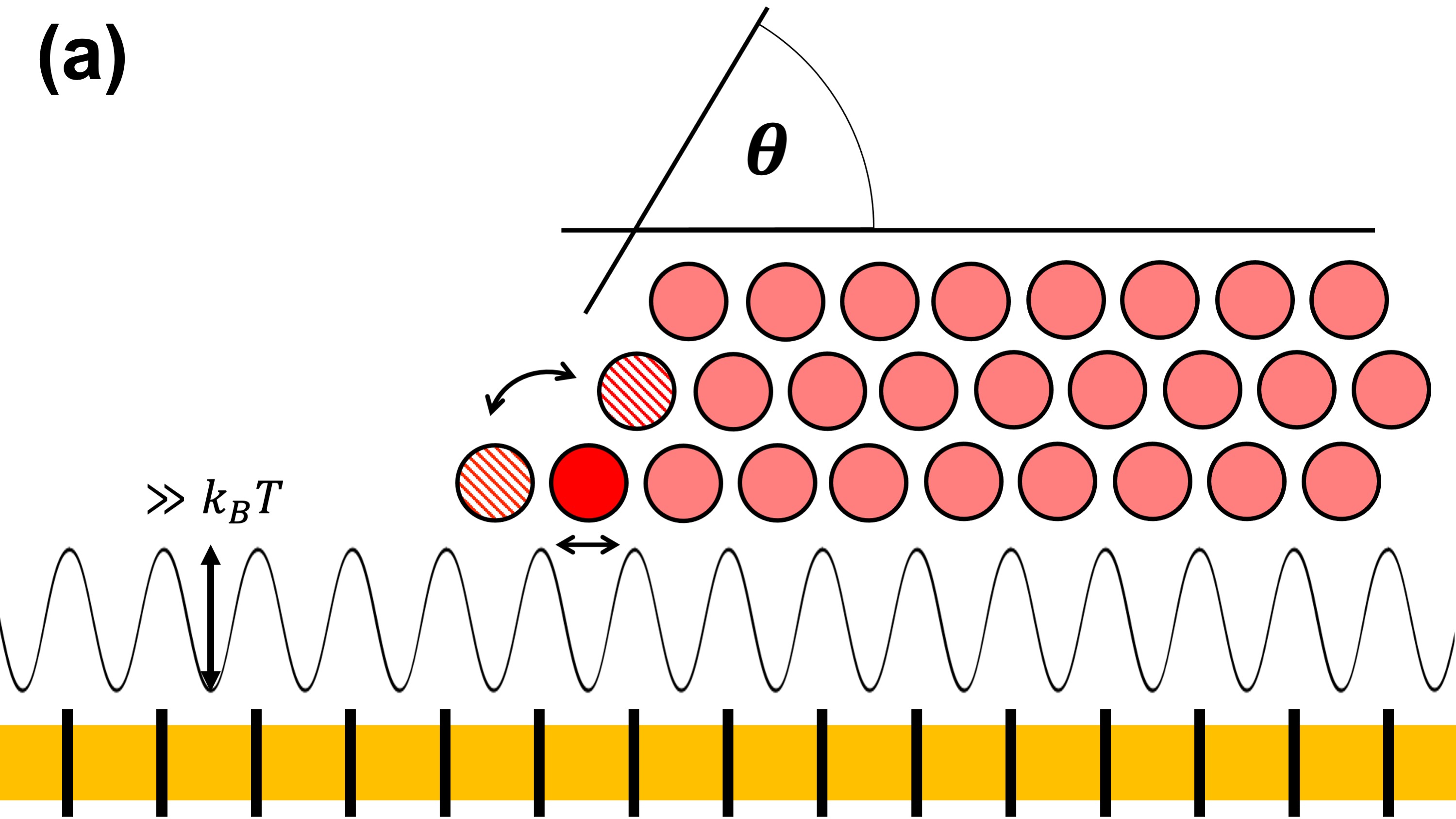}\\
    \vspace{0.5cm}
    \includegraphics[width=0.45\textwidth]{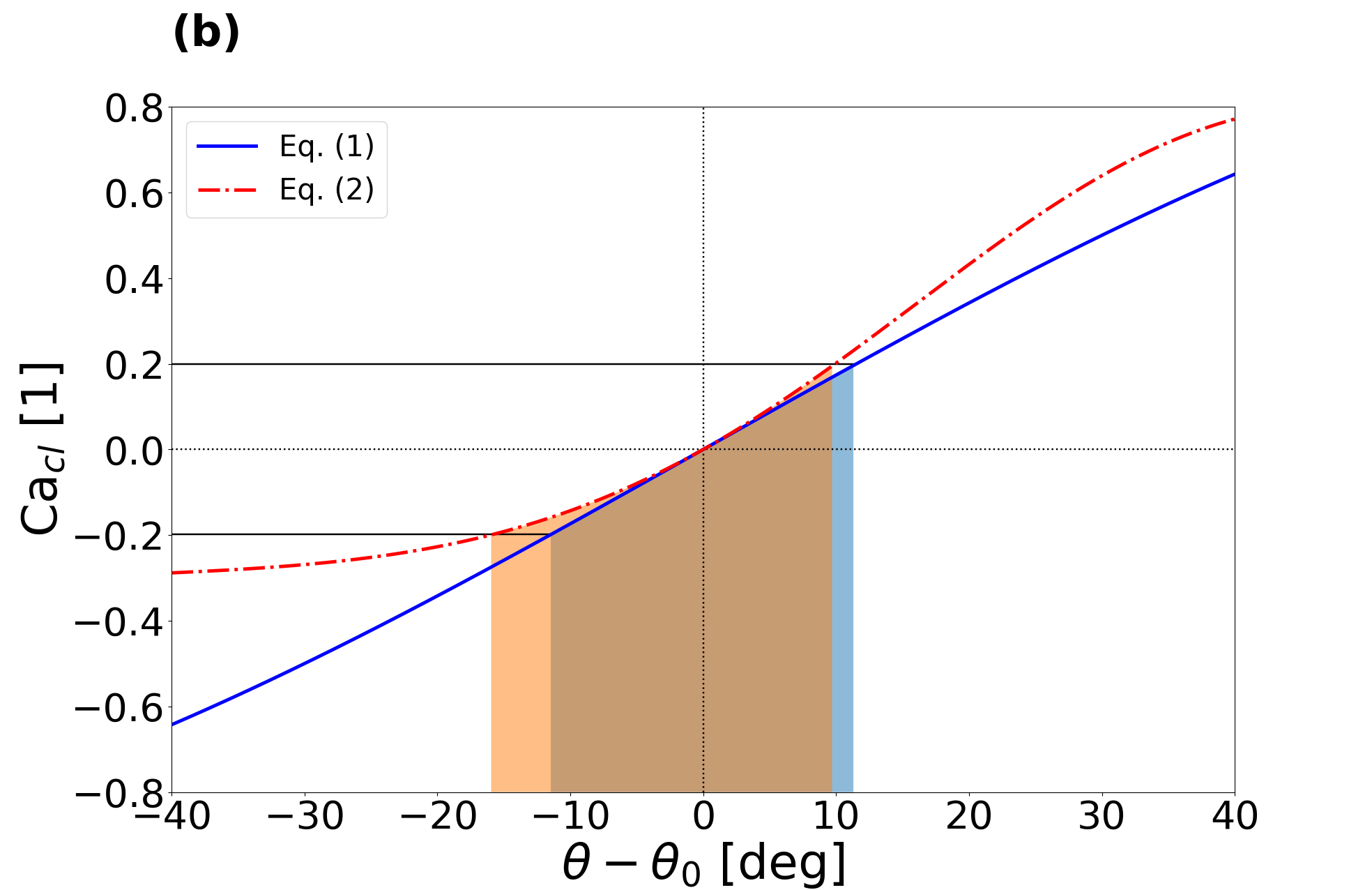}
    \caption{(a) Illustration of the situation captured by the two-layer model: molecules that are too close to the substrate rattle without escaping from potential wells while molecules from the fluid layer above can roll and jump over. (b) Exemplification of the asymmetry predicted by eq. \ref{eq:jump-roll}: for the same symmetric interval of contact line speed the interval of deviation from the equilibrium contact angle is skewed towards negative values; the contact line speed is re-scaled by $\gamma/\mu$, see equation \ref{eq:nondimensional} ($\theta_0=90^\circ$, $\mu_f=1=\overline{\mu}_f\cdot\exp(a/4)$ and $a=1.0$).}
    \label{fig:two-layers}
\end{figure}

Equation \ref{eq:linear-mkt} is symmetric, in the sense that it predicts a moving contact line to experience the same friction for the same absolute speed (or equivalently the same deviation from the equilibrium contact angle), regardless of whether the contact line is advancing (i.e. it moves towards the liquid phase) or receding (i.e. it moves towards the vapour phase). The same consideration holds for the full non-linear mobility expression. MKT is based on the assumption that molecules adsorbed in the potential minima of the substrate are mobile enough to effectively carry the contact line motion. If the surface is very hydrophilic, however, it may happen that energy barriers in this adsorption/desorption process are too high and it becomes energetically preferable to advance a contact line by transporting molecules from the above liquid layers instead (Fig.~\ref{fig:two-layers}a). Johansson and Hess \cite{johansson2019friction} included this phenomenology in the contact line friction model by envisioning an angle-dependent contact line friction parameter:
\begin{equation}    \label{eq:jump-roll}
    \mu_f(\theta) = \overline{\mu}_f \exp\{a\epsilon(\theta)\} \; .
\end{equation}
The reasoning behind equation \ref{eq:jump-roll} is that molecules in the second layer will find the harder to  `roll over' the stronger the interactions with neighbouring water molecules and the smaller the dynamic contact angle. Parameter $a$ captures the effect of the neighbouring liquid molecules; \com{$\epsilon(\theta)$ is a angle-dependent energy barrier which scales with distance needed for a molecule in the second fluid layer to roll over an adsorbed molecule in the first layer. Such distance will depend on the inclination angle between the two layers, that is the dynamic contact angle. Using simple geometric arguments, the barrier has been previously estimated by Johansson and Hess\cite{johansson2019friction} as: $\epsilon(\theta) = (0.5\sin\theta+\cos\theta)^2$}. This contact line friction model (which we will refer to as \textit{two-layer model}) not only incorporates a richer phenomenology compared to MKT, but also predicts situations where friction is asymmetric, that is the same contact line speed causes different deviation of the contact angle depending on the wetting direction (Fig.~\ref{fig:two-layers}b).

When discussing results, the following non-dimensional quantities for the contact line friction and the contact line speed will be used:
\begin{equation}    \label{eq:nondimensional}
    \overline{\mu}_f^* = \frac{\overline{\mu}_f}{\mu} \; , \quad \mbox{Ca}_{cl} = \frac{\mu U_{cl}}{\gamma} \quad ,
\end{equation}
being $\mu$ the viscosity of the fluid. Since $\mu_f$ and $\mu$ have the same physical dimensions, one can interpret $\mu_f^*$ in terms of `how strong' is solid-liquid friction compared to the friction between liquid layers. $\mbox{Ca}_{cl}$ can be regarded as the contact line capillary number. Superscripts will be dropped for the sake of readability. Advancing and receding contact lines will be implicitly distinguished by the sign of the contact line capillary number ($\mbox{Ca}_{cl}>0$: advancing, $\mbox{Ca}_{cl}<0$: receding).

\section{Simulation and analysis methods} \label{sec:mod-met}

\subsection{Non-equilibrium molecular dynamics simulations}

\begin{figure}[htbp]
    \centering
    \includegraphics[width=0.35\textwidth]{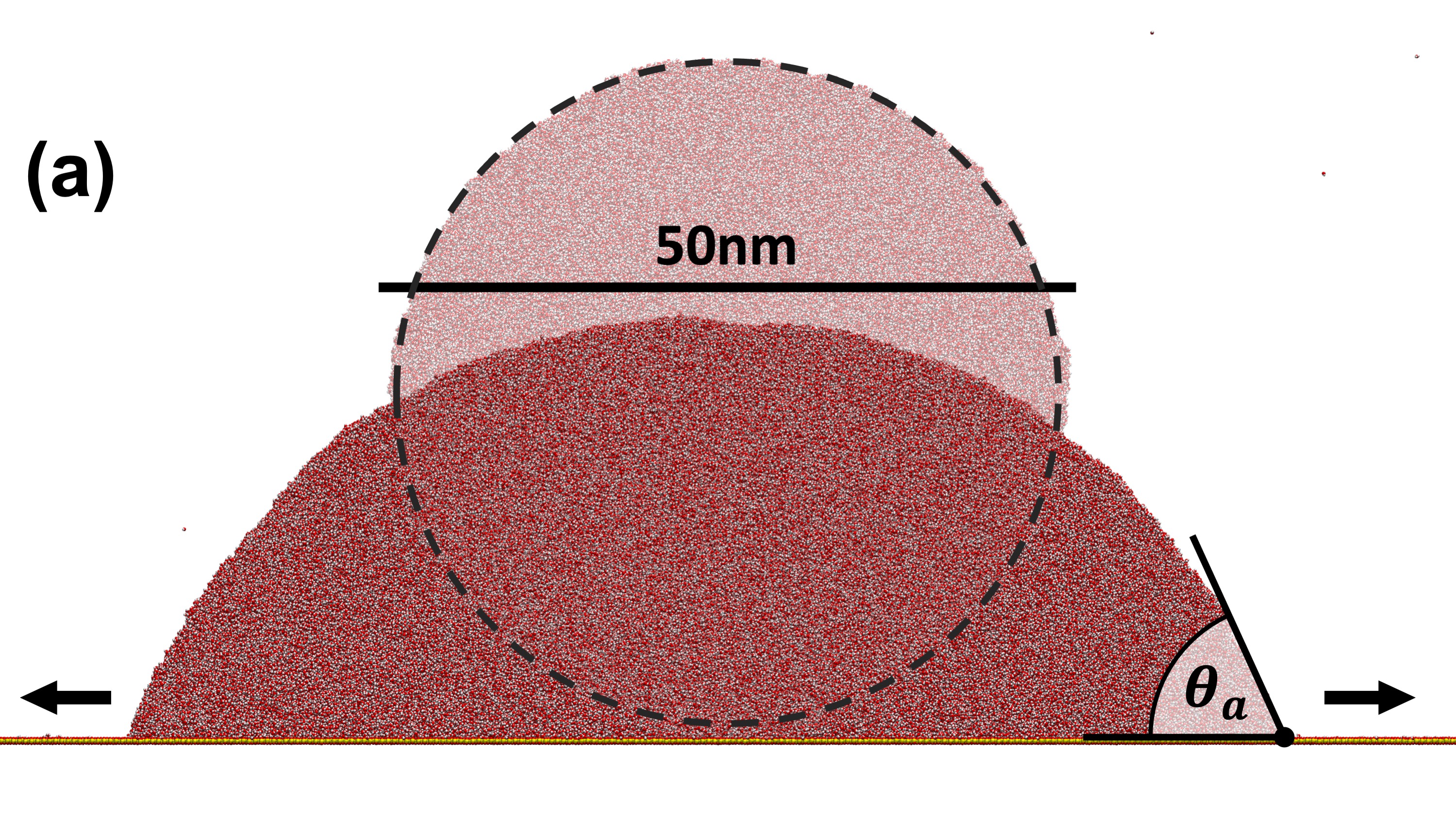}\\
    \vspace{0.5cm}
    \includegraphics[width=0.35\textwidth]{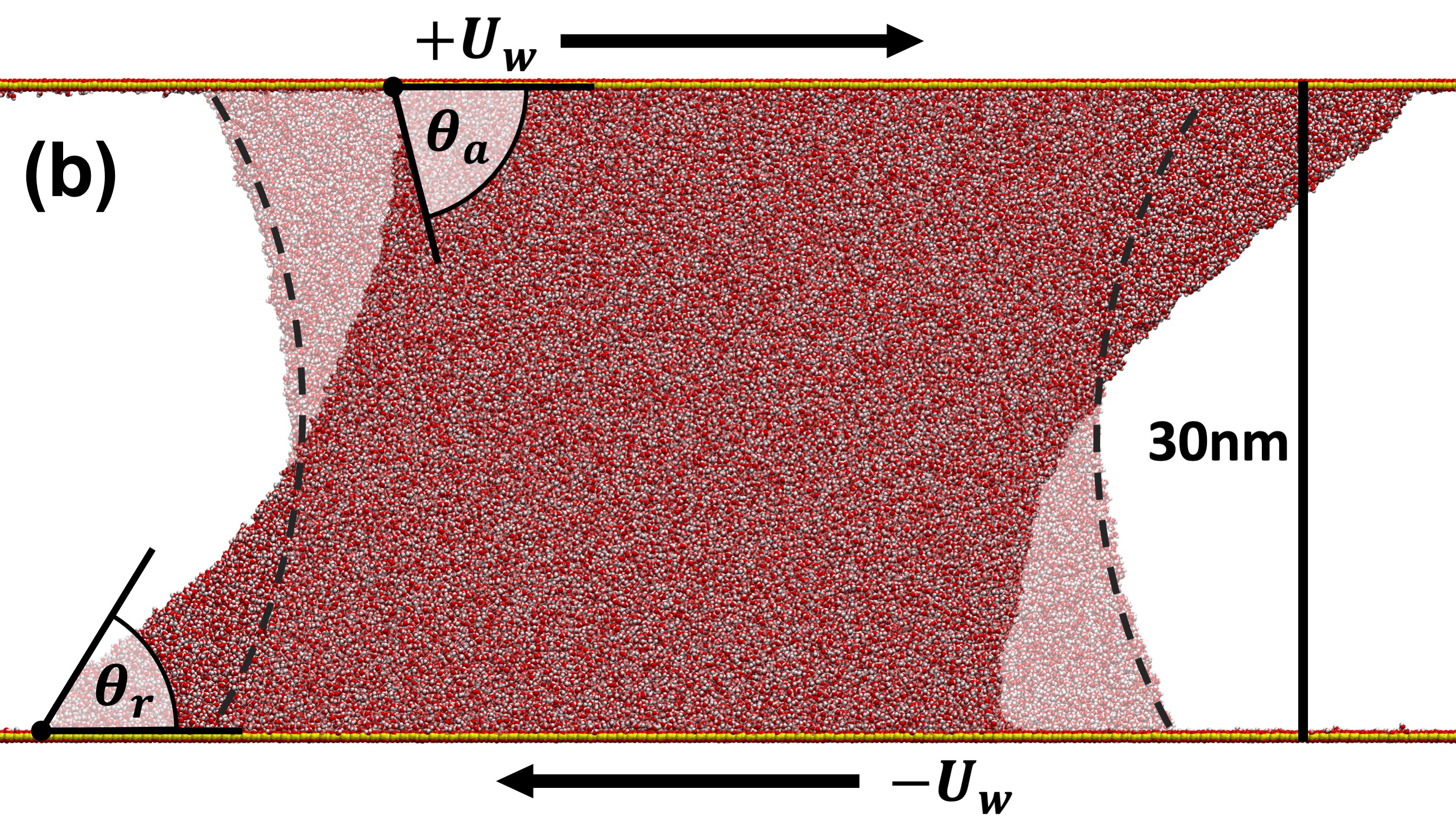}
    \caption{Illustration of the two wetting modes: spontaneous (a) and forced (b). The transparent figures on the background and the dashed lines correspond to the initial condition and the initial interface respectively. The subscripts for the contact angles `\textit{a}' and `\textit{r}' indicate whether the contact line is advancing or receding.}
    \label{fig:md-setups}
\end{figure}

\begin{figure*}[htbp]
    \centering
    \includegraphics[width=0.70\textwidth,trim={0 11.0cm 0 11.0cm},clip]{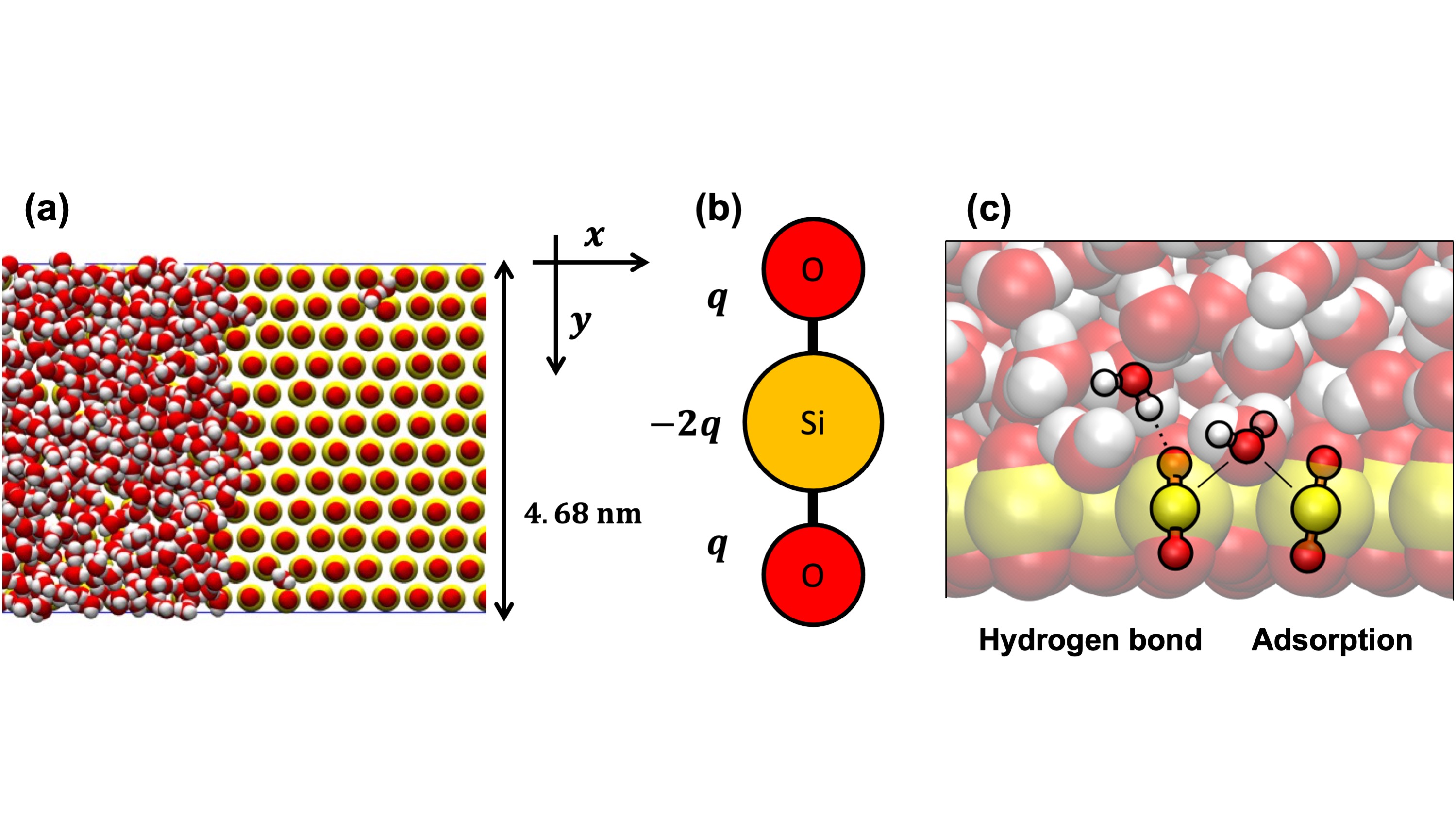}
    \caption{(a) Top-down view of a contact line ($\theta_0<90^\circ$). (b): Ball-and-stick model of the silica quadrupoles composing the solid substrate, with partial charges. (c) Illustration of the main hydrophilic interactions between water and silica. \com{Figure adapted from L\={a}cis et al.\cite{lacis2021pellegrino}}}
    \label{fig:md-details}
\end{figure*}

The molecular systems are simulated using GROMACS. The liquid quasi-2D droplets (Fig.~\ref{fig:md-setups}a) and menisci (Fig.~\ref{fig:md-setups}b) consists respectively of 300305 and 172933 SPC/E water molecules \cite{watermodels}. The substrates are composed of silica quadrupole molecules arranged in an hexagonal lattice (Fig.~\ref{fig:md-details}a). The substrate wettability can be tuned by changing the value of the partial charge $q$ on the oxygen atoms of silica (Fig.~\ref{fig:md-details}b). Albeit the lattice of silica quadrupoles is not physically realistic, the liquid-substrate combination is the simplest one reproducing the hydrophilic interaction that would characterise an high-friction surface  (Fig.~\ref{fig:md-details}c).

Non-equilibrium MD simulations are driven either by the initial condition being out of equilibrium or by a steady external action. In droplet spreading simulations a thin cylinder of SPC/E molecules is initialised so that the distance between the lower liquid-vapour interface and the upper oxygen atoms of the silica molecules is $\sim0.7$ nm. The droplet will spontaneously wet the surface until an equilibrium state characterised by the contact angle $\theta_0(q)$ is reached. Then, by increasing $q$ the droplet will spread even more over the surface (which is made more hydrophilic), while upon lowering $q$ the droplet will retract and de-wet the surface (which is made more hydrophobic). Changing the wettability of the surface thus allows to simulate both advancing and receding contact lines.

Shear droplet simulations, on the other hand, are driven by constantly moving the silica walls. The confined droplet is first initialised as a rectangular prism with height-to-width ratio $\sim0.75$ and let relax until it reaches its equilibrium meniscus shape. Then, the silica walls are moved in opposite directions with velocity $U_w$. If the wall velocity is below a critical velocity $U_{cr}(\theta_0)$, then the motion of the contact line is steady, i.e. the contact line velocity is zero if viewed from an absolute frame of reference or $-U_w$ if viewed from the frame of reference of the moving wall. Thus, the speed at which the contact line moves can be controlled. For $U_w\ge U_{cr}$ a critical transition occurs, which depending on the surface's wettability manifests either as a liquid film deposition at the receding contact line or as vapour entrainment at the advancing contact line.

\begin{table}
\centering

\caption{Equilibrium contact angle $\theta_0$ and critical capillary number $\mbox{Ca}_{cr}=\mu U_{cr}/\gamma$ as a function of the partial charge $q$ on silica oxygens.}
\begin{ruledtabular}
\begin{tabular}{lll}
    $q$ & $\theta_0$ & $\mbox{Ca}_{cr}$ \\
    \hline
    $-0.40e$ & $127^\circ$ &  $0.375\pm0.075$ \\
    $-0.60e$ & $95^\circ$ & $0.1375\pm0.0125$ \\
    $-0.67e$ & $69^\circ$ & $0.0625\pm0.0125$ \\
    $-0.74e$ & $38^\circ$ & $0.0125\pm0.00245$ \\
\end{tabular}
\end{ruledtabular}
\label{tab:equilibrium-critical}
\end{table}

The speed of the contact line will always be defined with respect to the inertial frame of reference moving alongside the corresponding substrate. Viscosity and surface tension for SCP/E water are the same as the ones in L\={a}cis et al. \cite{lacis2020johansson,lacis2021pellegrino}, that is $\gamma=5.78\times10^{-2}$ Pa$\cdot$m and $\mu=8.77\times10^{-4}$ Pa$\cdot$s. The critical capillary number for shear droplet configurations and the equilibrium contact angle as function of the partial charge $q$ are reported in Table~\ref{tab:equilibrium-critical}. Additional information on the parametrization of the force field can be found in the supplemental material.

\subsection{Interface and contact angle extraction}

Given the size of the molecular system, analysing fully-detailed and finely-sampled atomistic trajectories over scales of tens of nanoseconds would be prohibitive. The measurement of the contact line speed and contact angle does not require the knowledge the positions and velocities of all atoms at every time, but only of the position of the vapour-liquid interface; this one can be extracted as a contour line of the liquid density map $\rho(x,z)$. The density map is obtained by binning the atomic positions on-the-fly, that is concurrently with the running simulation, on a structured grid with spacing $\Delta x=\Delta z\simeq0.2$ nm. 

\begin{figure*}[!htb]
    \centering
    \includegraphics[width=0.70\textwidth,trim={0cm 0cm 0cm 0cm},clip]{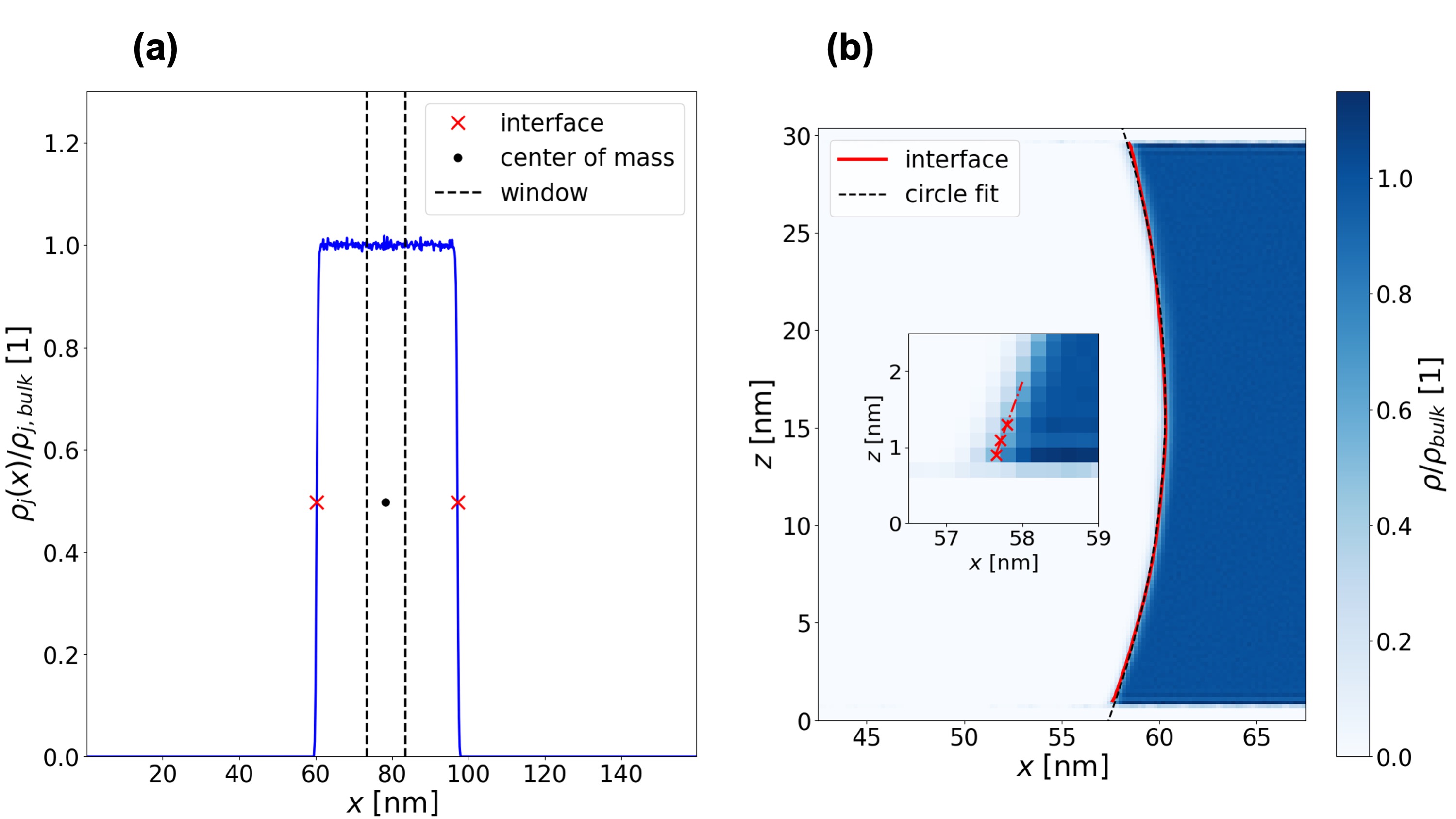}
    \caption{(a) Profile of a horizontal density slice showing the interface points and the averaging window used to define local bulk density. (b) Equilibrium interface profile fitted both with a circle and with linear interpolation (inset). Results for a confined meniscus with $\theta_0=69^\circ$.}
    \label{fig:interface}
\end{figure*}

The extraction of the interface is performed using a local half-density criterion: for each coordinate $z_i$ corresponding to the center of a bin the local density profile $\rho_j(x)$ is extracted; the local bulk density $\rho_{j,bulk}$ is defined as average of the density values in a region $\Delta x_{bulk}=10$nm, centered on the center of mass of the density slice \com{(Fig.~\ref{fig:interface}a)}. The interface point then is obtained by linearly interpolating the center of the two cells where the density is closer to $1/2\cdot\rho_{j,bulk}$. This approach produces a smooth interface curve despite the weak density layering occurring close to solid walls \com{(Fig.~\ref{fig:interface}b)}. Moreover, it is not required to measure the interface location for each $z_j$, but only for a few bins close to the solid substrate. Based on the analysis in \cite{lacis2021pellegrino}, the contact line location is identified with the interface point in the bin at $z_j=0.9$ nm (resp. $z_j=29.466$ nm for the upper wall in the shear droplet setup), which is about $0.3$ nm above the vertical coordinate of silicon atoms. 

In an equilibrium configuration one may identify the contact angle via the tangent to the best least-squares circle fit of the interface shape, following Young-Laplace law. The same approach obviously cannot be used when out of equilibrium; in such cases, which are the most interesting for this work, a linear interpolation of the first three interface points is used instead. Among all the interpolation techniques tried, this one provided more consistent results with the results in \cite{lacis2021pellegrino}.

Contact line velocity is not constant for spreading droplets. Obtaining the velocity from the signal of the contact line position over time via finite difference would yield extremely noisy results; hence a rational polynomial fit is performed on the position and the speed is obtained via analytical differentiation, as in  \cite{toledano2020hidden}. A numerical scanning on the number of polynomial coefficients is performed in order to single out the functional forms that produces the best fit for advancing and receding contact lines:

\begin{equation*}    
    x_{a}(t) = \frac{a_3 t^3 + a_2 t^2 + a_1 t + a_0}{a_{-1} t + 1} \; ,
\end{equation*}
\begin{equation}     \label{eq:rational-fit}
    x_{r}(t) = \frac{b_2 t^2 + b_1 t + b_0}{b_{-2} t^2 + b_{-1} t + 1} \; .
\end{equation}

The number of coefficients of the rational polynomial function is different depending if the contact line is advancing or receding, but it is kept the same for all equilibrium contact angles. It is important to remark that equations \ref{eq:rational-fit} carry no physical meaning and are employed only for filtering purposes.

\subsection{Analysis of near-wall conformation of water molecules}

In order to study the near-contact-line conformation and mobility or water molecules we analyse detailed molecular trajectories sampled every 1ps in a time window of 4ns. Since the full molecular system would be far too large to process, we select atomic positions from a control volume of size $l_x\simeq2.82$ nm $\times l_y\simeq4.68$ nm $\times l_z\simeq1.44$ nm centred at the moving contact line. In shear droplet simulations the contact line position is stationary in the frame of reference of the simulation box, making the selection easier. We focus on the case of $\theta_0=69^\circ$ \com{and $\theta_0=38^{\circ}$}, at equilibrium.

\begin{figure*}[htbp]
    \centering
    \includegraphics[width=0.35\textwidth,trim={0 0 7.5cm 0},clip]{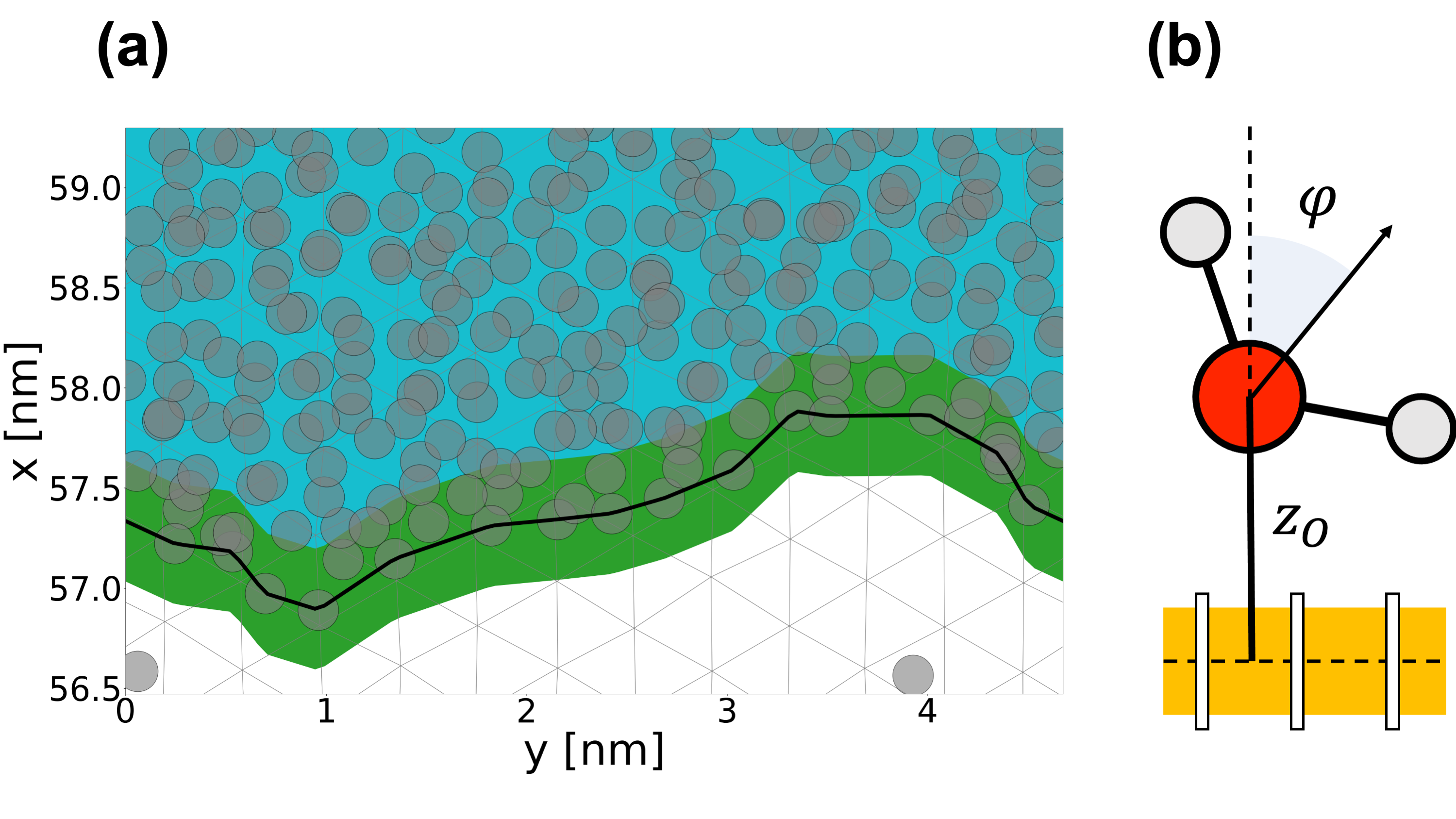}
    \hspace{25pt}
    \includegraphics[width=0.12\textwidth,trim={25cm 0 0 0},clip]{img/microscopic1-update.png}
    \caption{(a) Top-down view of the control volume used to investigate the conformation of water molecules. The green shaded region represents the contact line region, while the triangular mesh indicates the position of silica molecules (vertices). (b) Illustration of the degrees of freedom adopted to describe water molecules conformation.}
    \label{fig:microscopic}
\end{figure*}

Molecular coordinates are projected on the $(x,y)$ plane and a clustering procedure using DBSCAN algorithm is conducted to discriminate between the water molecules in the liquid wedge and the ones adsorbed in the substrate ahead of the contact line. Then, the contact line curve $x_{cl}(y)$ is approximated by the subset of the concave hull (alpha-shape) facing the solid-vapour interface (Fig.~\ref{fig:microscopic}a). The contact line molecules are defined as the ones having the x-coordinate of oxygen atoms within $d_{H_2O}=0.3166$ nm from the contact line: $|x_{O,i}-x_{cl}(y_{O,i})|<d_{H_2O}$. It is worth mentioning that this procedure assumes the molecules closer to the contact line to be also closer to the solid wall, hence it can be employed only if $\theta<90^\circ$.

The conformation of water molecules is defined via two degrees of freedom: $z_O$, the unsigned distance along $z$ between the coordinate of the oxygen atom and the coordinate of substrate's silicon atoms, and $\varphi$, the polar angle of the molecule's dipole w.r.t. the unit vector along $z$ (Fig.~\ref{fig:microscopic}b). The 2D histogram of the distribution of d.o.f. $\omega(z_O,\varphi)$ is computed using a simple binning procedure. The histogram is then transformed into an energy landscape via the Boltzmann formula: $E=-k_BT\log[\omega(z_O,\varphi)]$. The wells in the profile of $E$ represent the preferable conformations for water molecules.

\section{Results}    \label{sec:res-dis}

\subsection{Best fit of the two-layer mobility model}

\begin{figure}[htbp]
    \centering
    \includegraphics[width=0.49\textwidth,trim={3cm 0 3cm 0},clip]{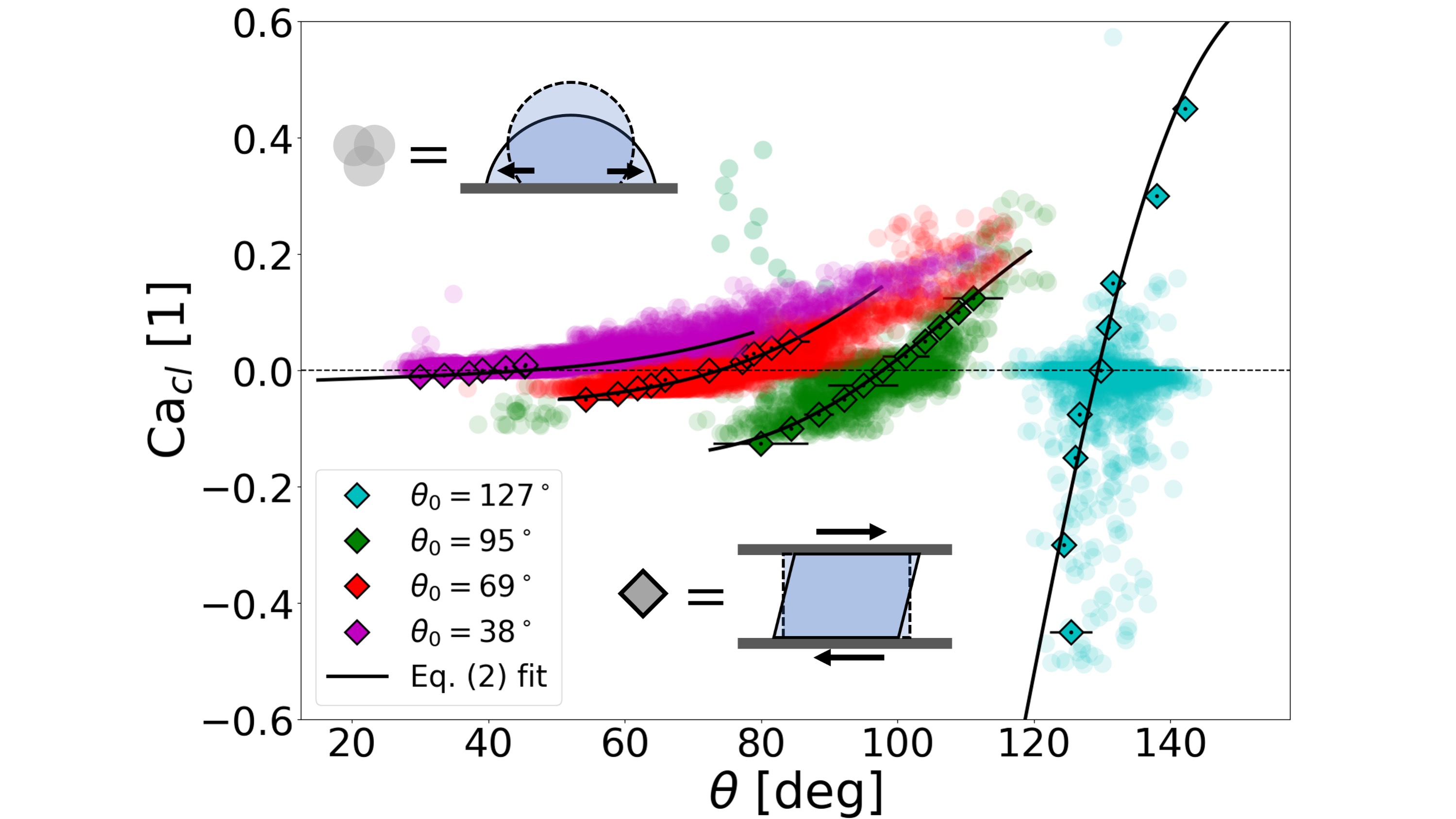}
    \caption{\com{Contact line capillary number against dynamic contact angle for different surface wettability and wetting modes; the solid black lines are the best fit of eq. \ref{eq:jump-roll}.}}
    \label{fig:model-fit}
\end{figure}

Figure \ref{fig:model-fit} displays the measurements of dynamic contact angle and contact line speed for spreading and shear droplet simulations. Despite the measurements from spreading droplet simulations showing significant variability, the quantitative agreement between forced and spontaneous wetting is evident for $\theta_0=95^\circ$ and $\theta_0=69^\circ$. In the case of $\theta_0=127^\circ$ and $\theta_0=38^\circ$ both wetting modes provide consistent results, albeit the range of contact line velocity is different; as a matter of fact, it is easier to collect measurements from droplet spreading simulations if the contact line is advancing when the substrate is hydrophilic (vice-versa if hydrophobic). The simulations therefore do not show any significant difference between forced and spontaneous wetting configuration in terms of contact angle - contact line speed relation. This is consistent with recent state-of-the-art molecular dynamics investigations \cite{toledano2020hidden} and corroborates the assumption that forms the bedrock of the contact line friction model, according to which the contact line mobility relation should be invariant on the wetting mode at the molecular scale.

\begin{table}[htbp]
\centering
\caption{Values and the uncertainty range of the fitting parameters for eq. \ref{eq:jump-roll}. The estimate for $a$ is the same for all values of $\theta_0$.}
\vspace{5pt}
$a=1.031\pm0.3$\\
\vspace{5pt}
\begin{ruledtabular}
\begin{tabular}{  c | r  r  r  r  }
    \hspace{0.1cm}$\theta_0$\hspace{0.1cm} & $127^\circ$ & $95^\circ$\hspace{0.1cm} & $69^\circ$\hspace{0.1cm} & $38^\circ$\hspace{0.1cm} \\
    \hspace{0.1cm}$\overline{\mu}_f$\hspace{0.1cm} & \hspace{0.1cm}0.255$\pm$0.013\hspace{0.1cm} & \hspace{0.1cm}1.72$\pm$0.074\hspace{0.1cm} & \hspace{0.1cm}2.39$\pm$0.39\hspace{0.1cm} & \hspace{0.1cm}5.89$\pm$0.69\hspace{0.1cm} \\
\end{tabular}
\end{ruledtabular}
\label{tab:fitting-parameters}
\end{table}

Once verified that both spontaneous and forced wetting modes produce consistent results, the next step is to fit the model described by equation \ref{eq:jump-roll}. In order to reduce the arbitrariness of the functional fit, the parameters $a$ and $\mu_f$ are fitted using separate criteria. In particular, since the model predicts $a$ to be dependent on the liquid composition but not on the surface's wettability, only one value will be fitted encompassing all the MD results. On the other hand, the contact line friction depends on the strength of the electrostatic interactions between water and silica, which in turns unequivocally expresses the equilibrium contact angle; hence, a different value of contact line friction will be fitted depending on the surface wettability $\mu_f(\theta_0)$. The obtained values of $a$ and $\mu_f(\theta_0)$ are shown in Table~\ref{tab:fitting-parameters} and show good agreement with the ones previously obtained in \cite{johansson2019friction}. The uncertainty range is estimated by performing 25 cross-validation steps where 20\% of the dataset is removed upon estimating the parameters. Additional details regarding uncertainty quantification are provided in the supplemental material.

The results in this section show the capability of the two-layer model to capture all four possible combinations of moving contact lines: forced and spontaneous, advancing and receding. In the following sections we portray the molecular physics supporting the model.

\subsection{Near-contact line conformation of water molecules}

\begin{figure}[htbp]
    \centering
    \includegraphics[width=0.45\textwidth,trim={0 0 0 0},clip]{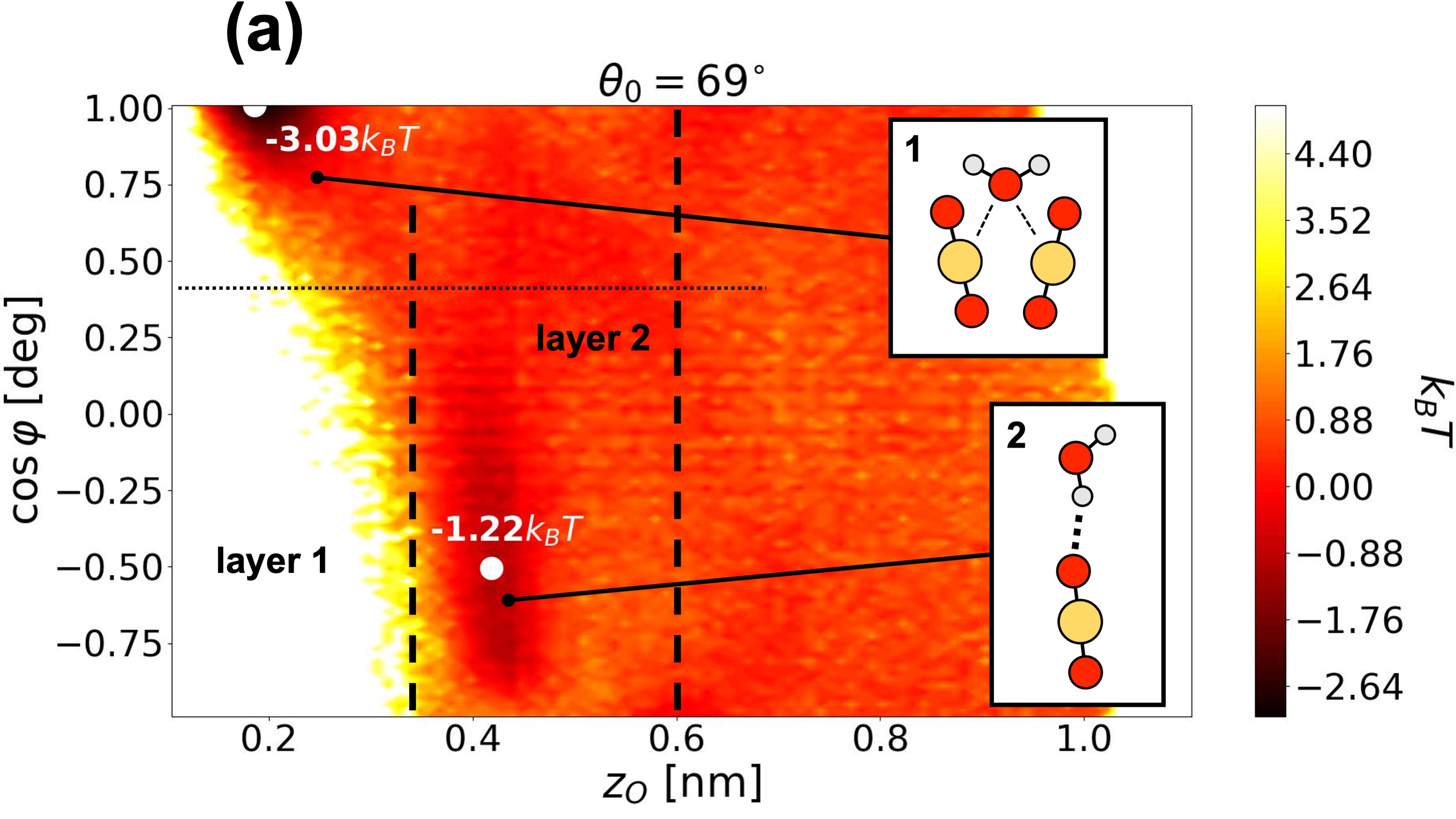}\\
    \vspace{0.5cm}
    \includegraphics[width=0.45\textwidth,trim={0 0 0 0},clip]{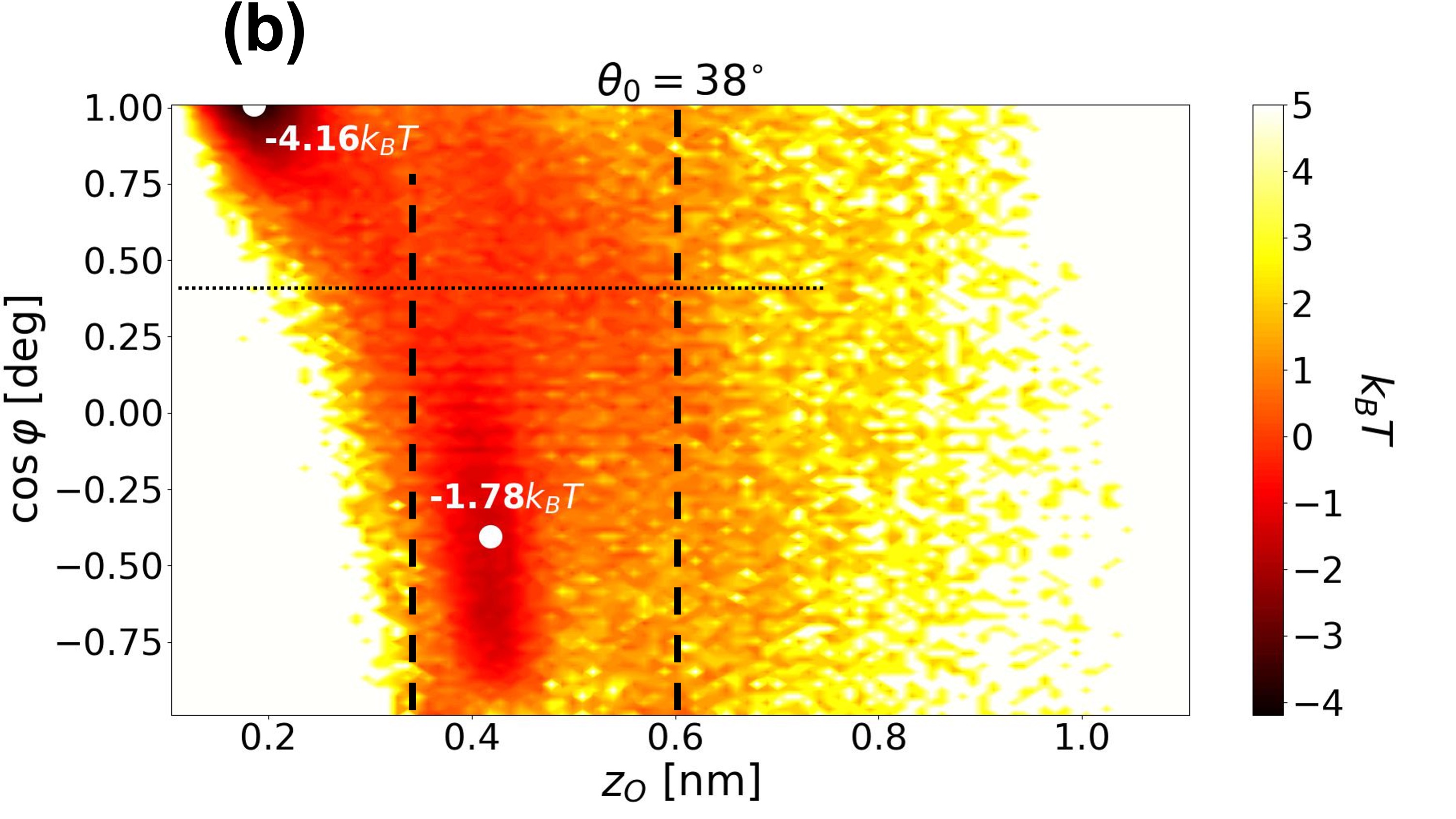}\\
    \vspace{0.5cm}
    \includegraphics[width=0.425\textwidth,trim={5cm 0 5cm 0},clip]{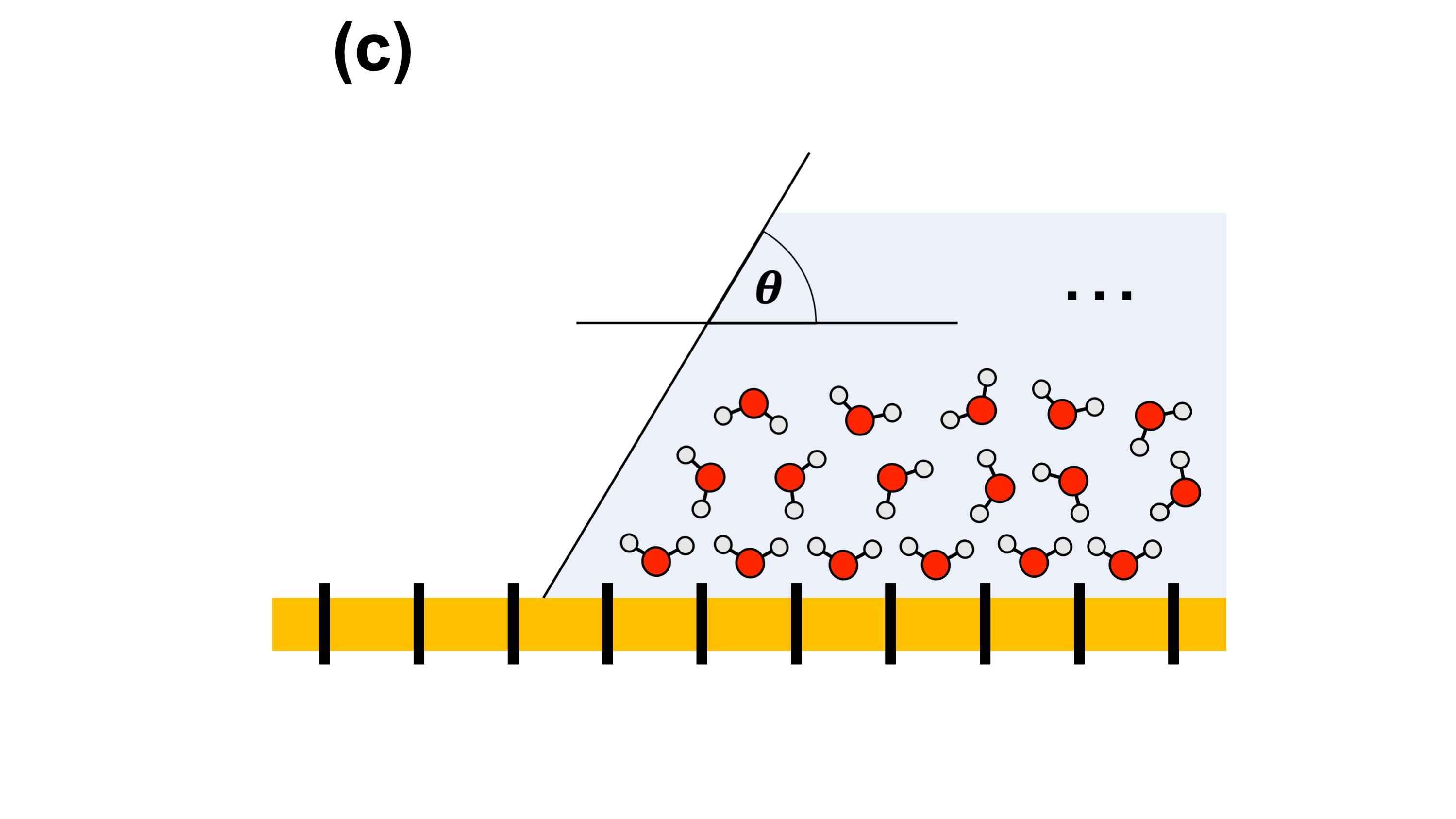}
    \caption{(a) Contour map of $E=-k_BT\log[\omega(z_O,\varphi)]$, that is the Boltzmann transformation of the distribution of the single-molecule d.o.f., for $\theta_0=69^{\circ}$. (b) Boltzmann transformation for $\theta_0=38^{\circ}$. (c) Representation of the qualitative differences in orientation and layering of near-wall water molecules.}
    \label{fig:conformation}
\end{figure}

Figure \ref{fig:conformation}a shows the contour lines of the energy landscape $E(z_O,\cos\varphi)$ obtained after the binning procedure, \com{for $\theta_0=69^{\circ}$}. Two wells can be spotted, the deeper having minimum at $(z_O\simeq0.2\mbox{ nm}, \cos\varphi\simeq1)$ and the other having minimum for $(z_O\simeq0.43\mbox{ nm}, \cos\varphi\simeq-0.63)$. While weak horizontal layering of water molecules has already been observed in \cite{lacis2021pellegrino}, it can be noticed that each water layer shows a preferential orientation of the electrostatic dipole. Water molecules in the wall-nearest layer have the dipole pointing upwards; their conformation thus corresponds to what has been labelled `adsorption' in Fig.~\ref{fig:md-details}c. Water molecules in the layer above have their dipole mainly pointing downward, although the distribution of $\cos\varphi$ is more outspread; their conformation corresponds to what has been labelled `hydrogen bonding'. \com{The landscape for $\theta_0=38^{\circ}$ (Fig.\ref{fig:conformation}b) appears qualitatively similar.}

The profile of $E(z_O,\cos\varphi)$ suggests that molecules in different layers of the fluid interact with the substrate in qualitatively different ways (Fig.~\ref{fig:conformation}b). It is therefore possible to speculate that molecules adsorbed in layer 1 would find it more difficult to escape the energy well compared to the ones in layer 2 due to being caged into the silica lattice. From $E(z_O,\cos\varphi)$ it is even possible to compute the energy barrier between layer 1 and 2 by taking the minimum free energy pathway. Using a simple Monte Carlo procedure, the energy barrier is roughly estimated as $\Delta E=1.31\pm0.014k_BT$; this value is not much outside the uncertainty range of parameter $a$, which, according to the discussion in section \ref{sec:line-friction}, scales the energy barrier associated with rolling motion. It must be pointed out that this barrier is modulated by the collective conformations of water molecules in the contact line region. Therefore, the value of the barrier computed from $E$ may overestimate the actual energy barrier for a molecular displacement event of a moving contact line, and thus should be considered only a zero-th order approximation. However, this approach to estimate displacement barriers is not dissimilar to the one taken by MKT, where single molecular jumps are considered instead of collective processes.

\com{

\section{Kinetics of contact line molecules}    \label{sec:kinematics}

\subsection{Modelling two-conformations kinematics}

In order to thoroughly validate the two-layer model, the kinematics of near-contact-line molecules needs to be studied. Reasoning in the framework of MKT, the speed of a contact line is dictated by the frequency of molecular displacement events and their distance. For the situation depicted in this work, the multicomponent molecular-kinetic approach by Liang et al.\cite{liang2010multicomponent} can prove instrumental in mapping kinematic information to different conformations of water molecules. Considering the contribution of `rolling' and `jumping' molecules to the displacement of the contact line alike the one of two different species, we can rewrite the speed of the contact line as:
\begin{equation}
    U_{cl} = \sum_{i=1}^2 \kappa_i \delta_i = \sum_{i=1}^2 \kappa^0_i\mbox{expt}_i\delta_i \; ,
\end{equation}
being $\kappa_i$ and $\kappa_i^0$ the non-equilibrium and the equilibrium displacement rates of species $i$, $\delta_i$ its displacement length and
\begin{equation}    \label{eq:tilting}
    \mbox{expt}_i=\exp\Big\{\frac{w_i\delta_i^2}{2k_BTx_i}\Big\}-\exp\Big\{-\frac{w_i\delta_i^2}{2k_BTx_i}\Big\}
\end{equation}
the exponential tilting of the potential energy landscape occurring when then contact line is out of equilibrium, with $w_i$ being the specific work of adhesion and $x_i$ the local mass fraction of species $i$. Ultimately, comparing the exponential tilting for each species should provide quantitative information about their relative effect to contact line mobility. However, a direct computation of expression \ref{eq:tilting} is unfeasible due to the difficulty in characterizing the work of adhesion. Conversely, assuming both $\kappa_i$ and $\kappa_i^0$ are quantifiable, one can obtain the tilting as $\mbox{expt}_i=\kappa_i/\kappa_i^0$.

\begin{figure}
    \centering
    \includegraphics[width=0.42\textwidth,trim={0 0 0 0 0},clip]{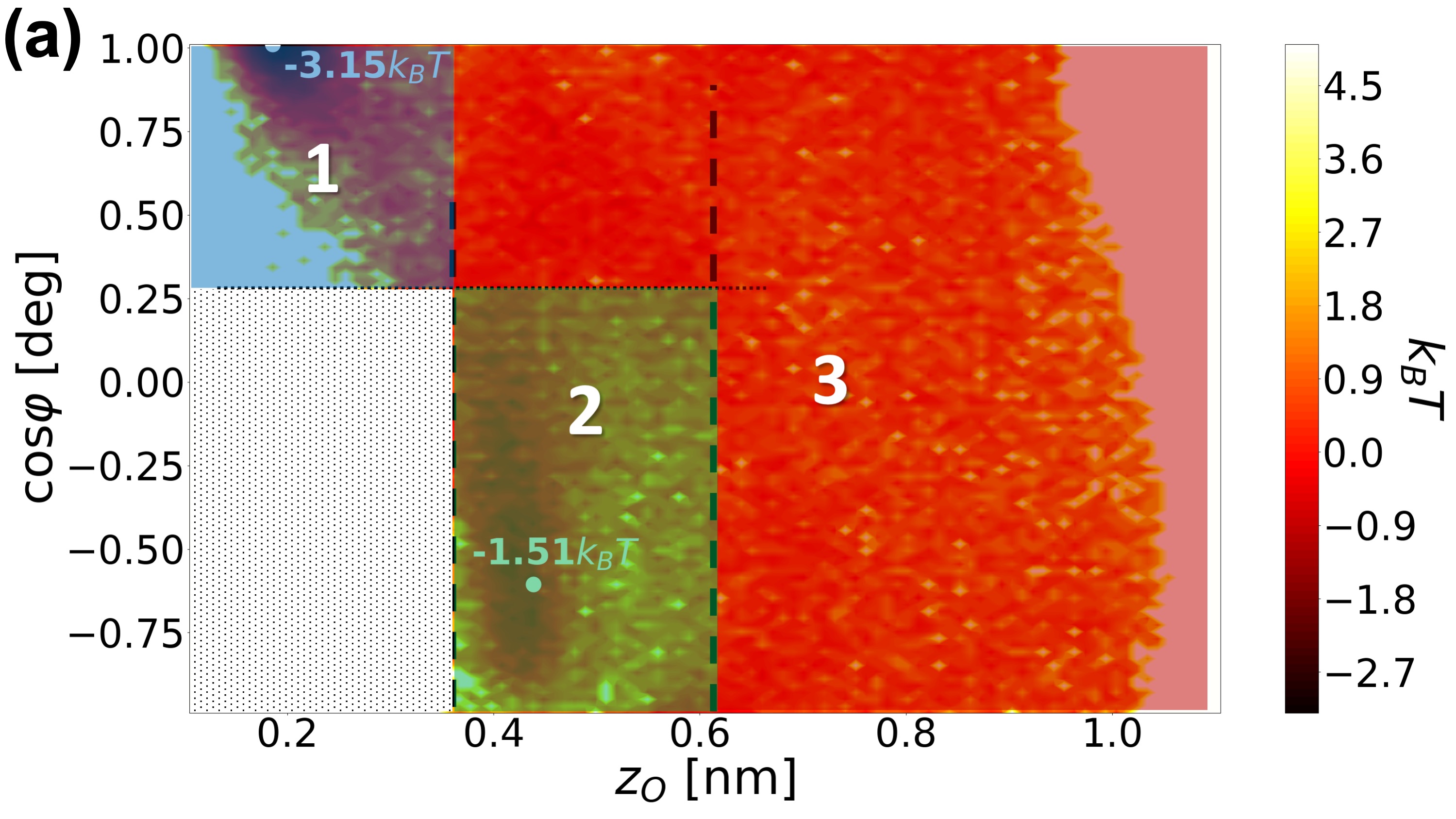}\\
    \vspace{0.5cm}
    \includegraphics[width=0.42\textwidth,trim={0 0 0 0},clip]{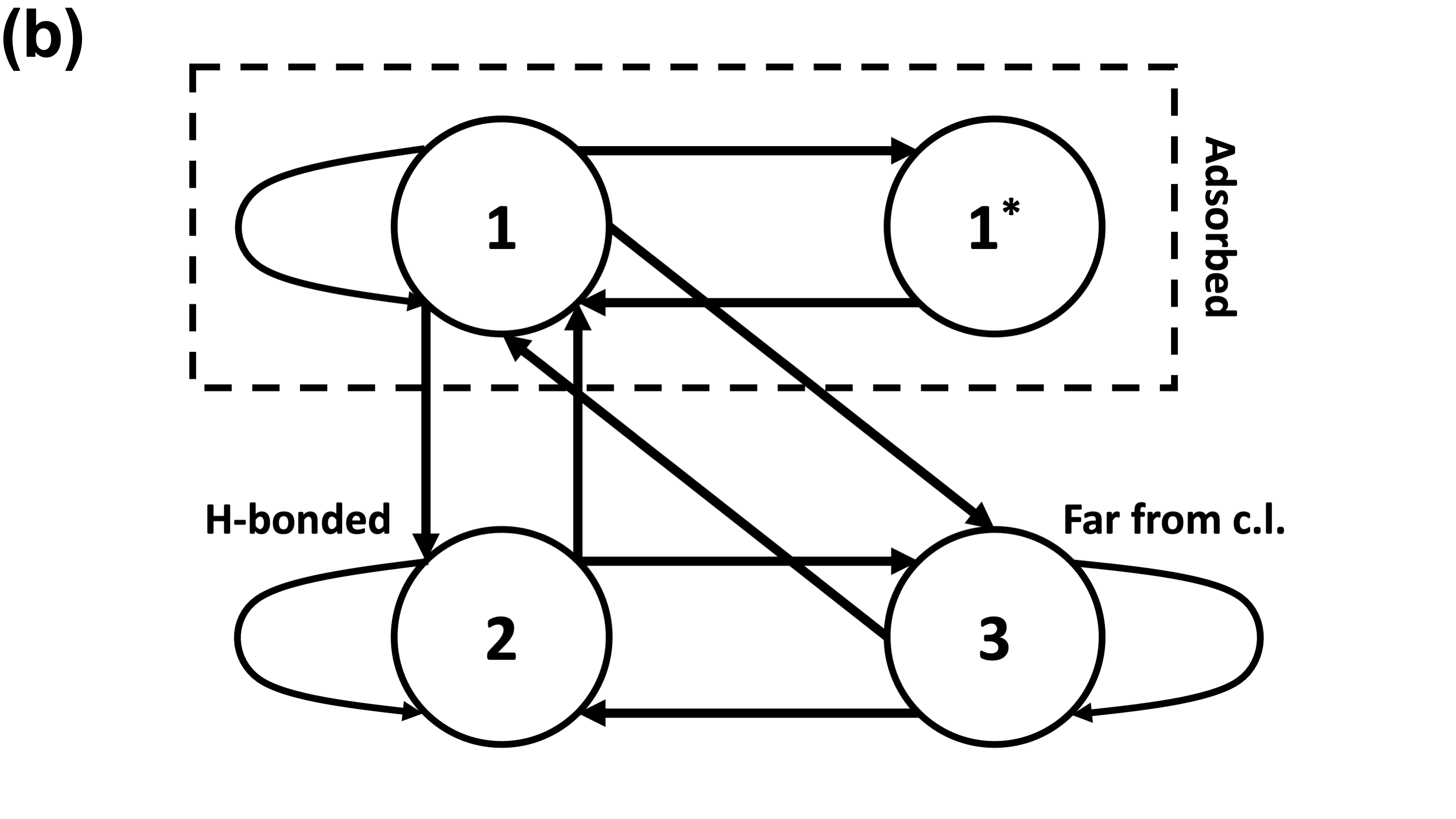}
    \caption{\com{(a) Labelling of the conformation states based on the coordinates $z_O$ and $\cos\varphi$. The separators between classes correspond to the lines $\cos\varphi\simeq0.25$, $z_O\simeq0.375$ and $z_O\simeq0.6$. (b) Schematics of the Markov chain used to model molecular motion in the contact line region.}}
    \label{fig:markov-model}
\end{figure}

The method we employ to extract the displacement rates from molecular simulation involves modelling the transition between different water conformations as a quasi-Markov process and compute the corresponding transition matrix. Let us label state \textbf{1} as the one of molecules adsorbed in the silica lattice, \textbf{2} as the one of molecule in the second layer that are hydrogen bonded and \textbf{3} the one of molecules that are away from the contact line, either in the bulk or at interfaces (Fig.\ref{fig:markov-model}a). A `rolling' event would correspond to the transition from \textbf{2} to \textbf{1}. In order to define a `jump' event we need to incorporate memory effects and define an additional state \textbf{1$^*$} which is employed only to describe adsorption to a different silica lattice cavity. Therefore the transition from \textbf{1} to \textbf{1$^*$} will be considered a `jump'. Being $p_{ij}(t_{lag})=p(i\rightarrow j|t_{lag})$ the conditional probability of a molecule in state \textbf{i} to transition to state \textbf{j} after time $t_{lag}$, the lag-dependent transition matrix is defined as:

\begin{equation}
    P_{ij}(t_{lag}) = p(i\rightarrow j|t_{lag}) \; .
\end{equation}

Fig.\ref{fig:markov-model}b presents a schematics of the resulting Markov chain. It is worth noticing that because of how \textbf{1$^*$} is defined, a transition to or from \textbf{1$^*$} by any other state except \textbf{1} is not possible, while $P_{1^*,1}(t_{lag})=1$, regardless of the lag time.

\subsection{Transitions probabilities and rates}

\begin{figure}
    \centering
    \includegraphics[width=0.475\textwidth,trim={0 0 2cm 0 0},clip]{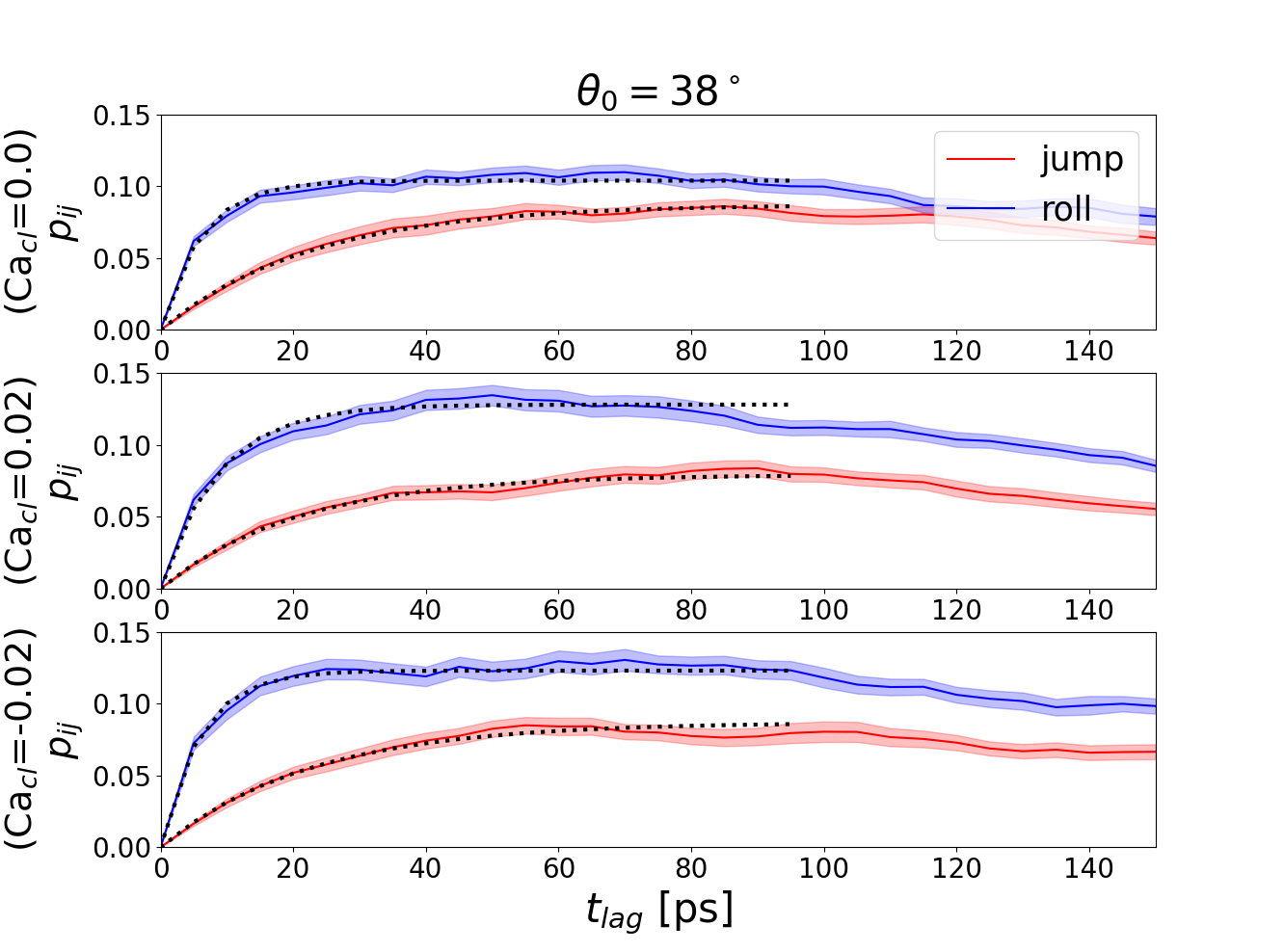}
    \caption{\com{Components of the transition matrix related to jumping and rolling motion. The shaded area represents the standard error on the transition probability for each time lag. The black dotted line is the best fit of the exponential saturation curve.}}
    \label{fig:transitions}
\end{figure}

The transition probability matrix is calculated for the case of $\theta_0=38^\circ$ at equilibrium and for $\mbox{Ca}_{cl}=\pm0.02$. The vector of time lags spans from 0ps to 150ps, sampled every 5ps. The transition rates for jumping and rolling (that is $P_{1,1^*}$ and $P_{2,1}$ respectively) are shown in Fig.\ref{fig:transitions}. We notice that contrary to what is usually observed in Markov models, the transition probability does not saturate, but rather starts to decrease after a maximum peak is reached. This behaviour is explained by the use of a fixed control volume to track the molecules in the contact line region, which may exit the volume after a sufficiently long time and thus cannot be tracked anymore.

\begin{table*}
    \caption{\com{Coefficients of the exponential saturation curves used to fit the transition probabilities of jumping and rolling over a span of lag times. The last row presents the exponential tilting of advancing and receding contact lines w.r.t. equilibrium. The uncertainty range is obtained via bootstrapping.}}
    \begin{ruledtabular}
    \begin{tabular}{c|r|r|r|r|r|r}
        \multirow{2}{*}{} &
        \multicolumn{2}{c|}{$\mbox{Ca}_{cl}=0.0$} &
        \multicolumn{2}{c|}{$\mbox{Ca}_{cl}=+0.02$} &
        \multicolumn{2}{c}{$\mbox{Ca}_{cl}=-0.02$} \\
        & Jump & Roll & Jump & Roll & Jump & Roll \\
        \hline
        $A$ [1] & 0.087$\pm$0.002 & 0.10$\pm$0.01 & 0.079$\pm$0.004 & 0.13$\pm$0.01 & 0.09$\pm$0.01 & 0.123$\pm$0.001 \\
        \hline
        $\tau$ [ps] & 22.6$\pm$1.6 & 6.13$\pm$3.57 & 20.6$\pm$2.5 & 8.75$\pm$3.45 & 22.4$\pm$2.6 & 5.94$\pm$0.68 \\
        \hline
        $\mbox{expt}$ [1] & - & - & 0.91$\pm$0.07 & 1.24$\pm$0.19 & 1.002$\pm$0.092 & 1.19$\pm$0.08
    \end{tabular}
    \end{ruledtabular}
    \label{tab:transitions}
\end{table*}

The transition probabilities are modelled with a simple exponential saturation curve $P_{ij}=A[1-\exp(-t/\tau)]$. A best least-squares fit of the curve is performed for time lags below the peak of the transition probability, which is obtained by direct visual inspection and corresponds roughly to 60ps for rolling transitions and 85ps for jumping transitions. Albeit using large fit times lead to a slight overestimate of $\tau$, they were taken this long to suppress noise. The estimates of $\tau$ and $A$ allow us to both inspect the relevant time scale of molecular motion and the c.l. displacement rate. Results are shown in table \ref{tab:transitions}. Roll events and jump events have two distinct time constants; the one of the former is comparable to the average lifetime of hydrogen bonds, whereas the one of the latter appears larger. The variability of time constants is not significant across capillary numbers, as all fall within the same range of uncertainty. The constants $A$ show less relative variability and can be used to estimate $P_{ij}$ for the value of $t_{lag}$ corresponding to their maximum. Furthermore, we assume that the ratio of displacement rates would scale as the ratio of transition probabilities:
\begin{equation}    \label{eq:exptilting}
    \mbox{expt}_i = \frac{\kappa_i}{\kappa^0_i} \sim \frac{P_{ij}}{P_{ij}^0} \; . 
\end{equation}
When estimating the exponential tilting using equation \ref{eq:exptilting} we obtain a higher value for the rolling population than for the jumping one, as reported in the last row of Tab.\ref{tab:transitions}. Hence, we can speculate that indeed the displacement of the contact line occurs mostly due to rolling between liquid layers rather than jumping on the silica lattice. However, when comparing advancing and receding contact lines, the uncertainty is too large to determine if there is a relative difference in the rolling-to-jumping ratios:
\begin{equation*}
    \mbox{expt}_{roll}^{(adv)}/\mbox{expt}_{jump}^{(adv)} = 1.36\pm0.31
\end{equation*}
\begin{equation}
    \mbox{expt}_{roll}^{(rec)}/\mbox{expt}_{jump}^{(rec)} = 1.19\pm0.19
\end{equation}
Hence, we can state that while the two-layer structure is important in modulating the mobility of the contact line, the effect on its asymmetry is verified only indirectly by fitting the model to observables such as the dynamic contact angle and the contact line speed.

}

\section{Discussion}   \label{sec:disc}

\subsection{Limitations of experimental observations}

As stated in the introduction, the disagreement found in experimental investigations of spontaneous and forced wetting has promoted the idea that a local description of dynamic contact lines, i.e. a functional relation between $U_{cl}$ and $\theta$, cannot exist \cite{blake1999nonlocal}. Our simulation results clearly show that there exists an agreement across wetting modes and that contact line friction well models the local mobility of contact lines. In this section we present some hypotheses on why experiments struggle to demystify dynamic wetting.

\com{The optical resolution scale accessible by conventional microscopy techniques (which will be referred to as $z_{res}$) is usually reported to be in the range of 500nm to 100$\mu$m.} \cite{blake1999nonlocal,duvivier2011viscosity,varagnolo2013strickslip,karim2016forcedspontaneous}. What can be directly measured in experiments is not the microscopic contact angle, but rather an apparent contact angle $\theta_{app}(z_{res})$ which results from the interface bending at the resolution length scale. This contact angle in general differs from the microscopic one. According to the classical description of viscous interface bending, the mismatch may be quantified using Cox's formula \cite{cox1986viscous,petrov1992combined}:
\begin{equation}    \label{eq:cox}
    \theta_{app}^3(z_{res}) - \theta^3 = + \mbox{Ca}_{cl}\log\big(z_{res}/\lambda\big) \; , 
\end{equation}
being $\lambda$ a microscopic cutoff length, which is often identified in the slip length. Although attempts to disentangle the effect of viscous bending from the one of contact line friction using experimental observation have been tried \cite{barriozhang2020friction}, the physical interpretation and the quantification of $\lambda$ still remain difficult, especially for fluid-substrate combinations showing negligible slip. Hence, the disentanglement often has to rely on an arbitrary parameter.

It is reasonable to model the dynamics as overdamped and friction-dominated for droplets or menisci with linear dimensions of $\sim10$ nm and for sufficiently small wetting speeds  \cite{doquang2015wetting}. The same cannot always be said for the length scale accessible by experiments, where inertial-capillary effects may start to become relevant. Cox's theory is based on the assumption that the dynamics is viscous-dominated, i.e. small Reynolds number; the picture changes for moderate Reynolds numbers \cite{varma2021inertial} and equation \ref{eq:cox} loses its validity. Inertial under-damped dynamics is not self-similar, hence experimental results are bound to differ unless the length scales, e.g. the droplet size, and the range of contact line velocities are both matched.

Because of the unavoidable limitations of experimental investigations, we wish to promote the idea of integrating molecular simulations in the validation process: an all-encompassing hydrodynamic model for moving contact lines is yet to be found and MD outputs could pave the way to its formulation. Already existing models can also greatly benefit from molecular benchmarks to perform parameter estimation, especially in the somewhat `pathological' case of no-slip surfaces. Following such intent, the results of the simulations conducted for this study are publicly accessible \cite{zenodoshear,zenodospread,zenodomicro}.


\subsection{Onset of wetting failure}   \label{sec:wetting-failure}

Because of the resolution limit described above, it would seem impossible to obtain experimental evidence to distinguish between different contact line friction models. Nevertheless, if experiments are inadequate to directly characterize the relation between contact line speed and contact angle, it is still possible to quantify the contact line speed at which wetting failure occurs \cite{snoeijer2006critical}. Wetting failure may be defined as the condition for which a steady moving contact line cannot exist; it is realized either by the deposition of a thin film at the receding contact line or by vapor entrainment at the advancing contact line \cite{snoeijer2013review}.

The onset of wetting failure is typically identified by the condition $\theta_{app}=0$, that is: the interface presents a zero-derivative inflection point at some distance from the contact line. It is possible to speculate whether the onset could be provoked by contact line friction. According to MKT, the maximum contact line speed is attained either for $\theta=0^\circ$ or $\theta=180^\circ$; assuming eq. \ref{eq:cox} is valid, it would imply $\theta_{app}=0^\circ$ or $180^\circ$ at any distance from the contact line. The two-layer model, on the other hand, predicts a maximum wetting speed attainable for some $\theta^*\in(0^\circ,180^\circ)$. The condition can be obtained by simply setting:
\begin{equation}
    \frac{\partial\mbox{Ca}_{cl}}{\partial\theta}\Bigg|_{\theta^*} = 0 \; .
\end{equation}

\begin{figure*}[htbp]
    \centering
    \includegraphics[width=0.675\textwidth,trim={0cm 0cm 0cm 0cm},clip]{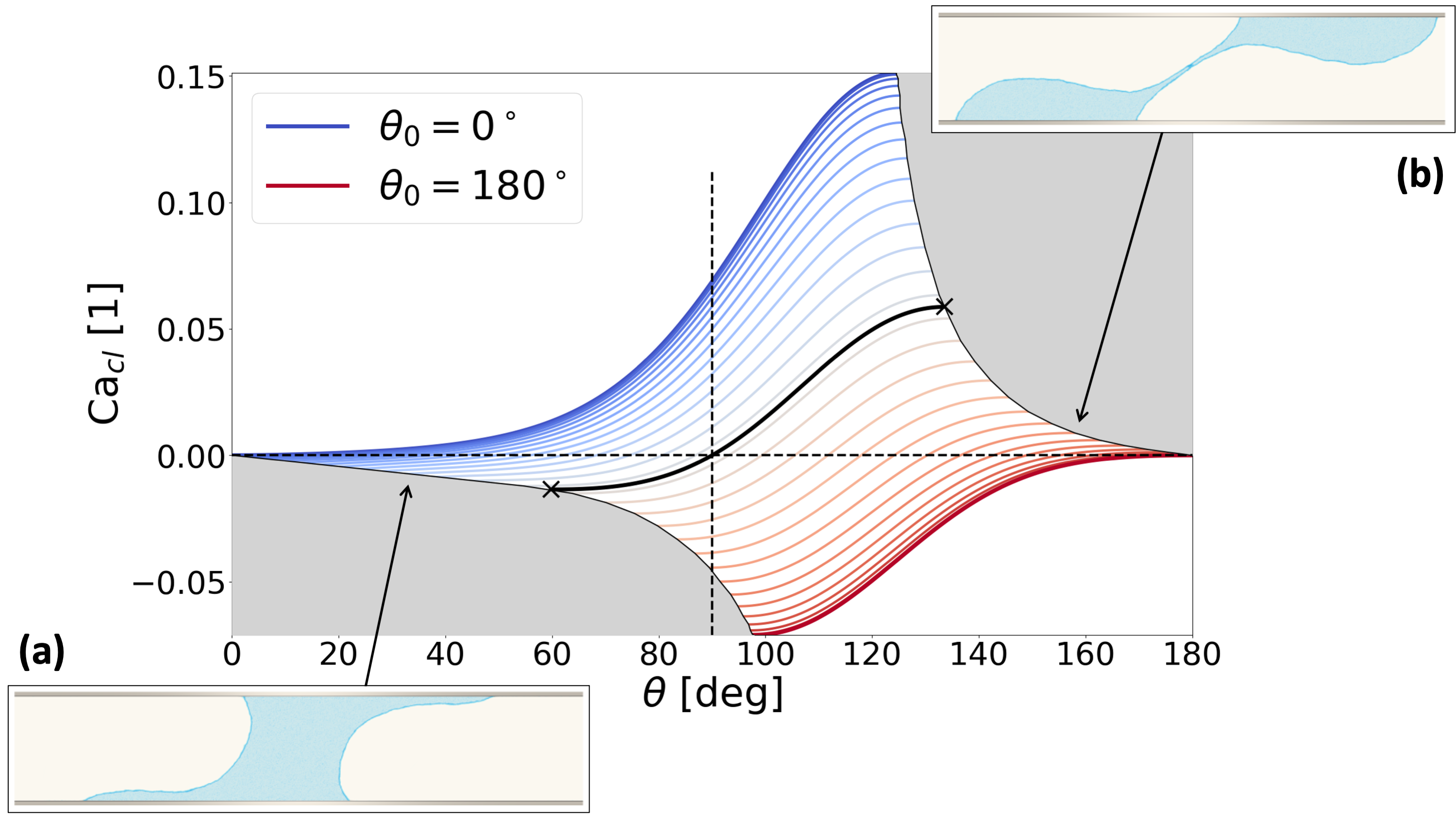}
    \caption{Example of profiles $(\mbox{Ca}_{cl},\theta)$, predicted by equation \ref{eq:jump-roll}. The color gradient correlates to $\theta_0$; $\overline{\mu}_f=5$, $a=1.5$. The shaded grey areas indicate the regions of instability for the moving contact line. The boundary on the left corresponds to the transition to film deposition (a), while the one on the right the transition to vapour entrainment (b). The profile highlighted by the solid black line corresponds to $\theta_0=90^\circ$; in this case $\theta_r^*\simeq59.78$ and $\theta_a^*\simeq133.58$.}
    \label{fig:max-speed}
\end{figure*}

Figure \ref{fig:max-speed} illustrates the contact line speed bounds consistent with equation \ref{eq:jump-roll}. Assuming $\theta^*$ is far enough from $0^\circ$ if the contact line is receding or $180^\circ$ if the contact line is advancing, it would be in principle possible to observe no zero-derivative inflection of the interface at the onset of wetting failure. Under such conditions, even macroscopic measurements of $\theta_{app}$ could suffice to discriminate between contact line friction models.

\section{Conclusions}   \label{sec:conc}

In this work we have expanded the application of the two-layer model by Johansson and Hess to cases of both advancing and receding contact lines, under spontaneous and forced wetting conditions, rationalising the contact line asymmetry observed in Lācis et al. The two-layer structure of liquid water close to the three-phase contact line has been illustrated. Our results illustrate the importance of the disruption of the order and orientation of liquid molecules caused by the interaction with the solid wall: this effect needs to be accounted for in order to model contact line friction in situations such as the one of water wetting silica-like surfaces. Models that over-simplify local molecular conformation may be insufficient to describe wetting/de-wetting asymmetry. \com{Baseline rates and probabilities for the rolling and hopping motions have been extracted from the transition time lags between the two identified conformations of water molecules in the contact line region. The results show the importance of the rolling of molecules in the second layers. However, unlike to the previously studied case of rapid spreading\cite{johansson2019friction}, the sub-critical capillary numbers attainable on a hydrophilic surface are too small to obtain a good signal-to-noise ratio for the transition rates and distinguish the exponential tilting on advancing contact lines from the one of receding contact lines.}

We have provided evidence to support the idea that the nanoscopic contact angle is a dynamic quantity and that its relation to contact line speed is invariant under wetting mode in a regime where the dynamics is overdamped. The resolution accessible by atomistic molecular dynamics simulation allows to quantify sub-continuous solid-liquid friction effects. It is important for experimental works incapable of resolving or inferring the curvature of liquid interfaces in the range of $\sim10$ nm from the moving contact lines to acknowledge this limitation. Nevertheless, studying the onset of wetting failure could help distinguish between alternative contact line friction models even on the basis of macroscopic experimental observations. From the point of view of continuous fluid dynamics modelling of moving contact lines, allowing the microscopic contact angle to deviate from its equilibrium value is paramount to reproduce the correct contact line physics, unless $\overline{\mu}_f\ll1$. Slip models that neglect contact line friction but manage to obtain quantitative agreement with MD or experiments should be regarded as effective, yet incomplete. 


We finally comment on potential ways to expand on the current investigation. Even though the silica surface reproduces the fundamental inter-molecular interactions with water, surfaces in reality are not atomistically flat and homogeneous \cite{quere2008review}. Topological and chemical defects or patterns can greatly influence contact line mobility; ways to adapt the contact line friction model to surfaces with defects or corrugations has been proposed in the past \cite{rolley2007cesium, lee2019topography}. 
In this work the contact line friction has been modified by tuning the interaction strength between the liquid and the substrate. Another interesting direction to explore would be to study the effect of changing the viscosity of the fluid \cite{duvivier2011viscosity}. 
Finally, it would be interesting to follow the thread of section \ref{sec:wetting-failure} and study how contact line friction affects the onset of wetting failure using molecular dynamics. This condition can be easily achieved by pushing the speed of solid walls beyond the critical droplet breaking value and observing contact line speed and interface curvature at the onset of instability.

\section*{Supplementary material}
See Supplemental Material at [URL will be inserted by publisher] for details of the force field used in molecular simulation and for a thorough illustration of uncertainty quantification for contact line friction parameters. 

\section*{Author Contributions}
\noindent\textbf{Michele Pellegrino}: Conceptualization, Methodology, Formal analysis, Investigation, Data Curation, Writing - Original Draft, Writing - Review \& Editing, Visualization.\\
\textbf{Berk Hess}: Conceptualization, Writing - Review \& Editing, Supervision, Project administration.

\section*{Author ORCID}
\noindent M. Pellegrino, \href{https://orcid.org/0000-0002-2603-8440}{orcid.org/0000-0002-2603-8440}\\
B. Hess, \href{https://orcid.org/0000-0002-7498-7763}{orcid.org/0000-0002-7498-7763}\\

\begin{acknowledgments}
    We acknowledge funding from Swedish Research Council (INTERFACE centre VR-2016-06119 and and grant nr. VR-2014-5680). We thank Dr. Petter Johansson for developing the code to perform data collection. We also thank Prof. Shervin Bagheri, Prof. Gustav Amberg, Prof. Stephane Zalenski, Dr. Ugis Lacis and Dr. Johan Sundin for the engaging scientific discussions which inspired the writing of this manuscript. Numerical simulations were performed on resources provided by the Swedish National Infrastructure for Computing (SNIC 2021/1-38) at PDC, Stockholm. Resources for data storage were provided by the Swedish National Infrastructure for Computing (SNIC 2022/23-46) and by Zenodo.
\end{acknowledgments}

\section*{Data Availability Statement}
The simulation output that supports the findings of this study is openly available in Zenodo at https://doi.org/10.5281/, reference number 6541983, 6445485 and 6385102. Analysis scripts in Python are available from the corresponding author upon reasonable request.

\section*{Conflict of Interest}
The authors have no conflicts to disclose.

\section*{References}
\bibliography{references} 

\end{document}